\documentclass[12pt,eqsecnum]{article}
\usepackage[dvips]{graphicx}
\usepackage{amssymb}
\usepackage{amsmath}
\usepackage{ulem}
\usepackage[dvipsnames]{xcolor}
\usepackage{soul}
\usepackage{ascmac}
\usepackage{cite}
\usepackage{url}
\makeatletter

\@addtoreset{equation}{section}
\makeatletter

\newcommand{\qed}{\hbox{\rule[-2pt]{6pt}{6pt}}}
\newcommand{\D}{{\rm d}}

\newtheorem{Prop}{Proposition}
\newtheorem{The}{Theorem}
\newtheorem{lm}{Lemma}
\newtheorem{dn}{Definition}
\newtheorem{Coro}{Corollary}

\newcommand{\dalm}{\kern1pt\vbox{\hrule height 0.9pt\hbox{\vrule width
0.9pt\hskip 2.5pt\vbox{\vskip 5.5pt}\hskip 3pt\vrule width 0.3pt}\hrule height
0.3pt}\kern1pt}

\begin{document}

\begin{titlepage}
\vfill
\begin{flushright}
\today
\end{flushright}

\vfill
%\vskip 1.0cm
\begin{center}
\baselineskip=16pt
{\Large\bf
Quest for realistic non-singular black-hole geometries: Regular-center type
}
\vskip 0.5cm
{\large {\sl }}
\vskip 10.mm
{\bf Hideki Maeda} \\

\vskip 1cm
{
Department of Electronics and Information Engineering, Hokkai-Gakuen University, Sapporo 062-8605, Japan.\\
\texttt{h-maeda@hgu.jp}

}
\vspace{6pt}
%\today
\end{center}
\vskip 0.2in
\par
\begin{center}
{\bf Abstract}
\end{center}
\begin{quote}
We propose seven criteria to single out physically reasonable non-singular black-hole models and adopt them to four different spherically symmetric models with a regular center and their rotating counterparts.
In general relativity, all such non-singular black holes are non-generic with a certain matter field including a class of nonlinear electromagnetic fields.
According to a criterion that the effective energy-momentum tensor should satisfy all the standard energy conditions in asymptotically flat regions, the well-known Bardeen and Hayward black holes are discarded.
In contrast, the Dymnikova and Fan-Wang black holes respect the dominant energy condition everywhere.
Although the rotating Fan-Wang black hole contains a curvature singularity, the rotating Dymnikova black hole is free from scalar polynomial curvature singularities and closed timelike curves.
In addition, the dominant energy condition is respected on and outside the event horizons in the latter case.
The absence of parallelly propagated curvature singularities remains an open question. 
\vfill
% \hrule width 5.cm
\vskip 2.mm
\end{quote}
\end{titlepage}

%<<<<<<<<<<<<< PACS NUMBER >>>>>>>>>>>>>>>%
%\pacs{
%04.20.Cv Fundamental problems and general formalism
%04.50.+h Gravity in more than four dimensions, Kaluza-Klein theory, unified field theories; alternative theories of gravity
%} 

% CECS-PHY-13/09

%\maketitle
%\section{}
%\subsection{}

\tableofcontents

\newpage

%======================================%
%<<<<<<<<<<<< SECTION 1 >>>>>>>>>>>>>>%
%======================================%
\section{Introduction}
\label{sec:intro}
As explicitly demonstrated in the Schwarzschild and Kerr vacuum solutions of the Einstein equations, the existence of a curvature singularity inside a black hole with a physically reasonable matter field is a well-known property in general relativity.
This fact implies that there appears an extremely curved spacetime region inside a black hole where quantum effects of gravity dominate and the classical Einstein equations are no longer valid.
Although a complete quantum theory of gravity is still unknown, it is widely believed that a regular description of the whole spacetime is possible in some quantum sense as a wave function of a hydrogen atom is regular at the classical central singularity in the Coulomb potential so that the Hamiltonian operator admits self-adjoint extensions~\cite{Bonneau:1999zq}.

Then, it might be possible that black holes without a singularity emerge in a classical theory realized in the low-energy limit of quantum gravity.
Such a theory should be a generalized theory of gravity which is non-minimally coupled with a scalar field and/or contains higher-curvature correction terms presumably.
Based on this philosophy, the first model of a {\it non-singular black hole} was proposed by Bardeen in 1968~\cite{Bardeen1968}.

If such non-singular black holes are realized in some generalized theory of gravity, the deviation of the theory from general relativity should be small in an asymptotically flat region far away from the event horizon.
Then, the effective energy-momentum tensor ${\tilde T}_{\mu\nu}$ is useful to sort out physically reasonable non-singular black-hole spacetimes without specifying a theory.
%----------------------- lemma ------------------------------%
\begin{dn}
\label{def:matter}
The {\it effective energy-momentum tensor} ${\tilde T}_{\mu\nu}$ is defined by ${\tilde T}_{\mu\nu}:=G_{\mu\nu}$ with units such that $c=8\pi G=1$.
\end{dn}
%----------------------- lemma ------------------------------%
In fact, various fundamental properties of black holes, such as the area theorem~\cite{Hawking:1971vc} or positive mass theorem for black holes~\cite{Gibbons:1982jg}, have been proven in general relativity under certain energy conditions without specifying a matter field.
In addition, it has been shown that a variety of matter fields are dynamically unstable if they violate the null energy condition, which is the weakest one among the standard energy conditions~\cite{Buniy:2005vh,Buniy:2006xf}.
Furthermore, the positive mass theorem asserts that the ADM mass is non-negative under the dominant energy condition and it is zero if and only if the spacetime is Minkowski~\cite{Schon:1979rg,Schon:1981vd,Nester:1982tr,Witten:1981mf}.
These results assert that the energy conditions prohibit pathological behaviors of spacetimes.

Therefore, it is natural to impose the following standard energy conditions~\cite{Hawking:1973uf,Maeda:2018hqu} on the effective energy-momentum tensor ${\tilde T}_{\mu\nu}$ at least in asymptotically flat regions:
\begin{itemize}
\item {Null} energy condition (NEC): ${\tilde T}_{\mu\nu} k^\mu k^\nu\ge 0$ for any null vector $k^\mu$.
\item {Weak} energy condition (WEC): ${\tilde T}_{\mu\nu} v^\mu v^\nu\ge 0$ for any timelike vector $v^\mu$.
\item {Dominant} energy condition (DEC): ${\tilde T}_{\mu\nu} v^\mu v^\nu\ge 0$ and $J_\mu J^\mu\le 0$ hold for any timelike vector $v^\mu$, where $J^\mu:=-{\tilde T}^\mu_{\phantom{\mu}\nu}v^\nu$. 
\item {Strong} energy condition (SEC): $\left({\tilde T}_{\mu\nu}-\frac{1}{2}{\tilde T}g_{\mu\nu}\right) v^\mu v^\nu\ge 0$ for any timelike vector $v^\mu$.
\end{itemize}
While the DEC implies the WEC, both of the WEC and SEC include the NEC as a limiting case.
Hence, all the standard energy conditions are violated if the NEC is violated.
In the singularity theorems~\cite{Penrose:1964wq,Hawking:1969sw}, the following geometric convergence conditions are used rather than the energy conditions for a matter field:
\begin{itemize}
\item {Null} convergence condition (NCC): $R_{\mu\nu} k^\mu k^\nu\ge 0$ for any null vector $k^\mu$.
\item {Timelike} convergence condition (TCC): $R_{\mu\nu} v^\mu v^\nu\ge 0$ for any timelike vector $v^\mu$.
\end{itemize}
The TCC includes the NCC as a limiting case and the Raychaudhuri equation implies that the gravitational force is attractive under these convergence conditions.
In general relativity, the NCC and NEC are equivalent in general, while the SEC is equivalent to the TCC in the absence of a cosmological constant $\Lambda$.
Hence, violation of the SEC implies that gravity becomes repulsive.
Such violation is not as serious as violations of other energy conditions as realized in the inflationary universe, for example.

In fact, it is known that the effective energy-momentum tensor ${\tilde T}_{\mu\nu}$ for the Bardeen black hole satisfies the WEC everywhere.
Nevertheless, it does not conflict with Penrose's singularity theorem~\cite{Penrose:1964wq} because the spacetime is not globally hyperbolic, so that one of the assumptions in the theorem is not satisfied.
However, as we will see, the DEC is violated in the asymptotically flat region in the Bardeen black-hole spacetime and therefore it should be discarded as a physically reasonable model of non-singular black holes.

In this context, the {\it limiting curvature condition} also plays an important role~\cite{Frolov:2016pav}.
\begin{itemize}
\item {Limiting curvature} condition (LCC): The curvature invariants are uniformly restricted by some universal value in the parameter space of the solution.
\end{itemize}
The LCC is based on the {\it limiting curvature principle} considered in~\cite{Markhov1982,Markov:1984ii,Polchinski:1989ae,Morgan:1990yy} and asserts that there exist a fundamental length scale $L$ so that ${\cal R}\le {\cal B}L^{-2}$ holds for any curvature invariant ${\cal R}$.
Here a dimensionless constant ${\cal B}$ may depend on curvature scalars but should not depend on solutions in the theory.
For example, suppose that a solution in a classical gravitation theory admits a regular de~Sitter (dS) or anti-de~Sitter (AdS) core such that
\begin{align}
\D s^2\simeq -(1-\lambda(m,l)r^2)\D t^2+\frac{\D r^2}{1-\lambda(m,l)r^2}+r^2(\D\theta^2+\sin^2\theta\D\phi^2) \label{LCC-center}
\end{align}
near $r=0$, where a constant $l$ with the dimension of length is made of coupling constants in the action and $m$ is a mass parameter as an integration constant.
The Ricci scalar $R$ near the regular center $r=0$ is given by the function $\lambda(m,l)$ of $l$ and $m$ as $R\simeq 12\lambda(m,l)$.
Then, the LCC requires that $|R|$, or equivalently $|\lambda(m,l)|$, should not be indefinitely large in the parameter space of $m\in (-\infty,\infty)$ because the effect of quantum gravity becomes significant and a classical description of spacetime is no more justified near $r=0$ for such $m$.
If the LCC is valid in a classical theory, the mass-inflation instability~\cite{Poisson:1989zz,Poisson:1990eh} of an inner horizon inside a non-singular black hole may be avoided in principle.
Such modified theories of gravity have been investigated in~\cite{Frolov:2016pav,Trodden:1993dm,Kunstatter:2015vxa,Chamseddine:2016ktu,Chamseddine:2019pux,Frolov:2021kcv}.

In fact, the Schwarzschild vacuum solution violates the LCC not only near the central singularity but also at any spacetime point if the ADM mass is sufficiently large.
In such a region, one should take higher-order correction terms into account in the action.
If there is a modified theory of gravity of which solutions satisfy the LCC, such correction terms are not necessary anywhere in the spacetime and the singularity resolution is completed within a classical framework.
Of course, there is a possibility that the LCC is never satisfied no matter how many correction terms are taken into account and then a full quantum theory of gravity is indispensable for a regular description of spacetime. 
In this sense, the LCC is more like a philosophy that singularity resolution is completed in a classical framework and its violation may not be a fatal problem.

Thus, from a conservative point of view, we propose the following seven criteria to single out physically reasonable non-singular black-hole spacetimes.

\begin{itembox}[l]{Seven criteria for physically reasonable non-singular black holes}
\begin{itemize}
\item {\bf C1:} Any kind of non-coordinate singularity is absent.

\item {\bf C2:} Closed causal curves are absent.

\item {\bf C3:} ${\tilde T}_{\mu\nu}$ satisfies the standard energy conditions in asymptotically flat regions.

\item {\bf C4:} ${\tilde T}_{\mu\nu}$ satisfies the standard energy conditions on the event horizon of a large black hole.

\item {\bf C5:} The limiting curvature condition (LCC) is respected.

\item {\bf C6:} Realized for a set of non-zero measure in the parameter space of the black-hole solution.

\item {\bf C7:} Dynamically stable.
\end{itemize}
\end{itembox}

\noindent
The criteria C1 and C2 are to avoid the singularity problem and causality problem, respectively.
The criterion C3 prohibits pathological behaviors of spacetime in asymptotically flat regions.
Similarly, the criterion C4 prohibits pathological behaviors of a ``almost-GR'' black hole with the sufficiently large area of the event horizon.
The criteria C1--C5 are purely geometrical and can be studied without specifying a theory.
In contrast, the criteria C6 and C7 are studied for non-singular black holes as solutions in a given theory.
The criterion C6 guarantees that singularities are absent without fine-tuning of integration constants in a solution representing a black hole so that a non-singular black hole is generic as a configuration of black hole.
In fact, as pointed out in a more general framework in~\cite{Chinaglia:2017uqd}, {\it all} the spherically symmetric non-singular black holes in general relativity obtained with a nonlinear electromagnetic field in the literature do not satisfy the criterion C6.
(See Appendix~\ref{sec:no-go-GR} for the proof in arbitrary $n(\ge 4)$ dimensions.)
As clearly observed in the magnetic solutions in~\cite{Fan:2016hvf}, without fine-tuning of the mass parameter, a black hole contains a curvature singularity at the center.

Among the above seven, the criterion C1 is the most fundamental one to single out black holes without singularities.
However, since the quantum effects of gravity dominate in extremely curved spacetime regions, only curvature singularities might be cured by them.
From this point of view, the criterion C1 may be weakened as follows.
\begin{itembox}[l]{Weak version of the criterion C1}
\begin{itemize}
\item {\bf W-C1:} Curvature singularities are absent.
\end{itemize}
\end{itembox}

\noindent
The criterion W-C1 permits conical singularities but not scalar polynomial curvature singularities and parallelly propagated (p.p.) curvature singularities~\cite{Hawking:1973uf}.
A conical singularity is classified as a {\it quasi-regular singularity}, which is not a curvature singularity but associated with peculiarities in the spacetime topology. (See section 6.1 in~\cite{Clarke:1994cw}.)

As we will review in Sec.~\ref{sec:history}, a large number of non-singular black-hole models have been proposed until now.
To single out physically reasonable models among them must be useful for astrophysical applications such as the gravitational lensing by black holes.
In the present paper, we study the geometrical criteria C1--C5 for non-singular black-hole spacetimes with a regular center.
In particular, the Bardeen black hole~\cite{Bardeen1968}, Hayward black hole~\cite{Hayward2006}, Dymnikova black hole~\cite{Dymnikova:2004zc}, and Fan-Wang black hole~\cite{Fan:2016hvf} are studied in detail.

The organization of the present paper is as follows.
In the next section, we will explain our metric assumptions and derive several general results.
We will also summarize the past research briefly there.
In Sec.~\ref{sec:RC}, we will study four different spherically symmetric non-singular black holes with a regular center and their rotating counterparts in detail.
Concluding remarks and future prospects will be given in the final section.
In Appendix~\ref{sec:no-go-GR}, we show that, under a set of assumptions, spherically symmetric non-singular black holes do not satisfy the criterion C6 in general relativity in arbitrary $n(\ge 4)$ dimensions.
In particular, a class of nonlinear electromagnetic fields is studied in detail.
In Appendix~\ref{sec:tidal}, we study tidal forces and Jacobi fields along ingoing radial timelike geodesics in the spherically symmetric spacetime (\ref{metric-app}).
Our conventions for curvature tensors are $[\nabla _\rho ,\nabla_\sigma]V^\mu ={R^\mu }_{\nu\rho\sigma}V^\nu$ and $R_{\mu \nu }={R^\rho }_{\mu \rho \nu }$.
The signature of the Minkowski spacetime is $(-,+,+,+)$, and Greek indices run over all spacetime indices.
We adopt units such that $c=8\pi G=1$.

%======================================%
%<<<<<<<<<<<< SECTION 1 >>>>>>>>>>>>>>%
%======================================%
\section{Black-hole geometries with a regular center}
\label{sec:metric}

Among several different types of non-singular black holes, we focus on the type with a regular center such as the Bardeen black hole in the present paper.
In this section, we give descriptions of such non-singular black-hole geometries and present several basic properties.
After that, we briefly summarize the research history of this type of non-singular black holes.

\subsection{Descriptions}

For almost all the non-singular black-hole spacetimes in the literature, the effective energy-momentum tensor ${\tilde T}_{\mu\nu}$ is of type I in the Hawking-Ellis classification~\cite{Hawking:1973uf,Maeda:2018hqu}.
The canonical form of the type-I energy-momentum tensor in the local Lorentz frame is given by 
\begin{equation} 
\label{T-typeI}
{\tilde T}^{(a)(b)}:={\tilde T}^{\mu\nu} {E}_\mu^{(a)} {E}_\nu^{(b)}=\mbox{diag}(\rho,p_1,p_2,p_3),
\end{equation}
where $\{{E}^\mu_{(a)}\}~(a=0,1,2,3)$ is a set of orthonormal basis vectors in the local Lorentz frame satisfying ${E}^\mu_{(a)}{E}_{(b)\mu}=\eta_{(a)(b)}$.
Here $\eta_{(a)(b)}$ is a Minkowski metric in the local Lorentz frame and the spacetime metric $g_{\mu\nu}$ is given by $g_{\mu\nu}=\eta_{(a)(b)}E^{(a)}_{\mu}E^{(b)}_{\nu}$.
Equivalent expressions to the standard energy conditions for a type-I matter field are
\begin{align}
\mbox{NEC}:&~~\rho+p_i\ge 0,\label{NEC}\\
\mbox{WEC}:&~~\rho\ge 0\mbox{~in addition to NEC},\label{WEC}\\
\mbox{DEC}:&~~\rho-p_i\ge 0\mbox{~in addition to WEC},\label{DEC}\\
\mbox{SEC}:&~~\rho+\mbox{$\sum_{j=1}^{3}$}p_j\ge 0\mbox{~in addition to NEC}\label{SEC}
\end{align}
for all $i(=1,2,3)$.
%We note that $p_2=p_3$ holds for spherically symmetric spacetimes.

\subsubsection{Non-rotating case}

In the spherically symmetric case, we consider non-singular black holes described by
\begin{align}
\label{metric-app}
\begin{aligned}
\D s^2=&-f(r)\D t^2+\frac{\D r^2}{f(r)}+r^2(\D \theta^2+\sin^2\theta\D\phi^2),\\
&f(r):=1-\frac{2M(r)}{r},
\end{aligned}
\end{align} 
where $M(r)$ is a mass function to be specified and the domain of $r$ is $r\in [0,\infty)$.
We note that this form of the metric is a strong assumption.
Although the most general spherically symmetric vacuum solution is given in the above form in general relativity by Birkhoff's theorem, there may be spherically symmetric vacuum solutions with the areal radius $\sqrt{g_{\theta\theta}}$ being a more general function of $r$ in modified theories of gravity or in the presence of matter fields.

In the spacetime (\ref{metric-app}), a trapped (untrapped) region in this spacetime is defined by $f(r)<(>)0$.
A regular null hypersurface $r=r_{\rm h}$ satisfying $f(r_{\rm h})=0$ is a Killing horizon associated with a Killing vector $\xi^\mu=(1,0,0,0)$, of which squared norm is $\xi_\mu\xi^\mu=-f(r)$.
A Killing horizon is referred to as {\it outer} if $\D f/\D r|_{r=r_{\rm h}}>0$, {\it inner} if $\D f/\D r|_{r=r_{\rm h}}<0$, and {\it degenerate} if $\D f/\D r|_{r=r_{\rm h}}=0$.
For an asymptotically-flat black hole, an event horizon is identical to the outermost outer Killing horizon.

The mass function $M(r)$ is chosen such that the spacetime is regular at $r=0$ and asymptotically flat as $r\to \infty$.
The following proposition gives a sufficient condition for a regular center.
%----------------------- lemma ------------------------------%
\begin{Prop}
\label{Prop:center-regularity}
Suppose that $M(r)$ in the spacetime (\ref{metric-app}) is expanded around $r=0$ as $M(r)\simeq M_0 r^{3+\alpha}$ with $M_0\ne 0$.
Then, $r=0$ is regular if and only if $\alpha\ge 0$.
In particular, as $r\to 0$, the spacetime is asymptotically locally flat for $\alpha>0$ and asymptotically locally dS (AdS) for $\alpha=0$ with $M_0>(<)0$. 
\end{Prop}
{\it Proof:}
Around $r=0$, the spacetime (\ref{metric-app}) and its Kretschmann invariant $K:=R_{\mu\nu\rho\sigma}R^{\mu\nu\rho\sigma}$ behave as
\begin{align}
\D s^2\simeq &-(1-2M_0r^{2+\alpha})\D t^2+\frac{\D r^2}{1-2M_0r^{2+\alpha}}+r^2(\D \theta^2+\sin^2\theta\D\phi^2),\label{metric-app-r=0}\\
K\simeq& 4M_0^2(\alpha^4 + 6\alpha^3 + 17\alpha^2 + 28\alpha + 24)r^{2\alpha}.\label{K-r=0}
\end{align}
Since the coefficient in Eq.~(\ref{K-r=0}) is non-zero for any value of $\alpha$, $K$ blows up as $r\to 0$ for $\alpha<0$, so that $r=0$ is a curvature singularity.
For $\alpha=0$, we obtain $\lim_{r\to 0}R^{\mu\nu}_{~~~\rho\sigma}=2M_0(\delta^\mu_{\rho}\delta^\nu_{\sigma}-\delta^\mu_{\sigma}\delta^\nu_{\rho})$ and hence the spacetime is locally dS (AdS) with $M_0>(<)0$ as $r\to 0$.
For $\alpha>0$, we obtain $\lim_{r\to 0}R^{\mu\nu}_{~~~\rho\sigma}=0$ and hence the spacetime is locally flat.
\qed
%----------------------- lemma ------------------------------%

As shown in Appendix~\ref{sec:tidal}, for $\alpha\ge 0$, tidal forces and Jacobi fields along an ingoing radial timelike geodesic $\gamma$ also remain finite as $r\to 0$.
For $-2<\alpha<0$, while tidal forces blow up as $r\to 0$ along $\gamma$ in general, at least two Jacobi fields diverge as $r\to 0$ along $\gamma$ with a particular value of its energy $E$.

\begin{figure}[htbp]
\begin{center}
%\rotatebox{-90}{
\includegraphics[width=0.8\linewidth]{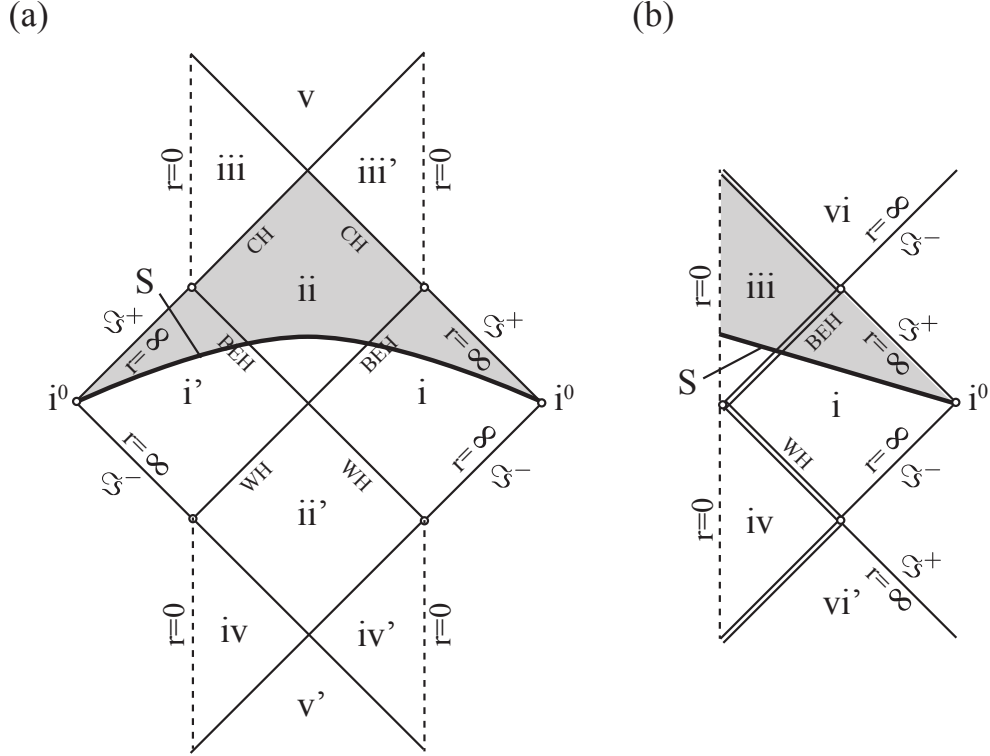}
%\subfigure[]{\includegraphics[width=0.7\linewidth]{Roberts-lambda1.eps}}
%\subfigure[]{\includegraphics[width=0.7\linewidth]{Roberts-lambda2.eps}}
%}
\caption{\label{PenroseSphericalNonsingularBH} Penrose diagrams of the Bardeen black hole with (a) two non-degenerate horizons and (b) one degenerate horizon.
The symbols $\Im^{+(-)}$ and $i^0$ stand for the future (past) null infinity and spacelike infinity, respectively. 
BEH, CH, and WH stand for the black-hole event horizon, Cauchy horizon (as an inner horizon), and white-hole horizon, respectively.
A double line in (b) represents a degenerate horizon.
}
\end{center}
\end{figure}

Since the metric function $f(r)$ converges to $1$ at a regular center $r=0$ and in an asymptotically flat region $r\to \infty$, the spacetime (\ref{metric-app}) is static in both regions.
This shows that the number of non-degenerate Killing horizons of a non-singular black hole with a regular center is always even.
The Bardeen black hole admits two Killing horizons at most depending on the parameters and its Penrose diagrams are drawn in Fig.~\ref{PenroseSphericalNonsingularBH}.
%------------<fig>---------------------------
%--------------<fig>-----------------------

As we will see in the next section, the effective energy-momentum tensor ${\tilde T}_{\mu\nu}$ for this type of spherically symmetric non-singular black holes can satisfy the DEC everywhere.
This fact does not conflict with Penrose's singularity theorem~\cite{Penrose:1964wq} under the following assumptions:
\begin{enumerate}
\item The spacetime is globally hyperbolic and admits a non-compact Cauchy surface S,

\item The NCC is satisfied, or equivalently ${\tilde T}_{\mu\nu}(:=G_{\mu\nu})$ satisfies the NEC, and 

\item S contains a trapped surface.
\end{enumerate}
In fact, the assumption 1 is violated in the spacetimes in Fig.~\ref{PenroseSphericalNonsingularBH}.
It is clear that the future domain of dependence $D^+({\rm S})$ (a shaded region) of a non-compact spacelike hypersurface S containing trapped surfaces does not cover the whole future of S.
This implies that S is just a partial Cauchy surface and the spacetime is not globally hyperbolic.
As a consequence, there is a Cauchy horizon (CH) as a part of the future boundary of $D^+({\rm S})$.

According to the criteria C3 and C4, we are going to check the energy conditions for ${\tilde T}_{\mu\nu}$ of the spacetime (\ref{metric-app}).
A natural set of orthonormal basis one-forms $\{E^{(a)}_\mu\}~(a=0,1,2,3)$ in the spacetime (\ref{metric-app}) consists of
\begin{align}
&E^{(0)}_\mu\D x^\mu = \left\{
\begin{array}{ll}
-\sqrt{f(r)}\D t & (\mbox{if}~f(r)>0)\\
-\sqrt{-f(r)^{-1}}\D r & (\mbox{if}~f(r)<0)
\end{array}
\right.,\\
&E^{(1)}_\mu\D x^\mu = \left\{
\begin{array}{ll}
-\sqrt{f(r)^{-1}}\D r & (\mbox{if}~f(r)>0)\\
-\sqrt{-f(r)}\D t & (\mbox{if}~f(r)<0)
\end{array}
\right.,\\
&E^{(2)}_\mu\D x^\mu=r\D \theta,\qquad E^{(3)}_\mu\D x^\mu=r\sin\theta\D\phi,
\end{align}
which give $\eta^{(a)(b)}=g^{\mu\nu}{E}_\mu^{(a)} {E}_\nu^{(b)}=\mbox{diag}(-1,1,1,1)$.
Then, the orthonormal components of ${\tilde T}_{\mu\nu}$ are given by Eq.~(\ref{T-typeI}) with 
\begin{align}
\rho=-p_1=\frac{2M'}{r^{2}},\qquad p_2=p_3=-\frac{M''}{r}\label{matter-nonrot}
\end{align}
independent of the sign of $f(r)$.
Throughout the present paper, a prime denotes differentiation with respect to $r$.
Equation~(\ref{matter-nonrot}) gives
\begin{align}
\label{matter-nonrotating}
\begin{aligned}
&\rho+p_1=0,\qquad \rho-p_1=\frac{4M'}{r^{2}},\\
&\rho+p_2=\rho+p_3=\frac{2M'-rM''}{r^2},\\
&\rho-p_2=\rho-p_3=\frac{2M'+rM''}{r^2},\\
&\rho+p_1+p_2+p_3=-\frac{2M''}{r}.
\end{aligned}
\end{align}
In the coordinate system (\ref{metric-app}), a Killing horizon defined by $f(r_{\rm h})=0$ is a coordinate singularity.
Nevertheless, by Proposition~3 in~\cite{Maeda:2021ukk} (or by Lemma~\ref{lm:horizon-matter-rotating} below with $a=0$), the standard energy conditions on a Killing horizon can also be studied by Eqs.~(\ref{matter-nonrot}) and (\ref{matter-nonrotating}) evaluated at $r=r_{\rm h}$.

Now we present a simple criterion to check the energy conditions around $r=0$.
%----------------------- lemma ------------------------------%
\begin{Prop}
\label{Prop:no-go-nonrot}
Suppose that $M(r)$ in the spacetime (\ref{metric-app}) is expanded around a regular center $r=0$ as 
\begin{align}
M(r)\simeq M_0r^{3+\alpha}+M_1r^{3+\alpha+\beta} \label{M-exp-r=0}
\end{align}
with $\alpha\ge 0$, $\beta> 0$, $M_0\ne 0$, and $M_1\ne 0$.
Then, the standard energy conditions are respected and violated around $r=0$ as shown in Table~\ref{table:EC-r=0}.
\begin{table*}[htb]
\begin{center}
\caption{\label{table:EC-r=0} The energy conditions near $r=0$ in the spacetime (\ref{metric-app}) with Eq.(\ref{M-exp-r=0}).}
\scalebox{1.0}{
\begin{tabular}{|c||c|c|c|c|c|}
\hline \hline
& {\rm NEC} & {\rm WEC} & {\rm DEC} & {\rm SEC} \\\hline
$\alpha=0, M_1>0$ & $\times$ & $\times$ & $\times$ & $\times$ \\ \hline
$\alpha=0, M_0>0, M_1<0$ & \checkmark & \checkmark & \checkmark & $\times$ \\ \hline
$\alpha=0, M_0<0, M_1<0$ & \checkmark & $\times$ & $\times$ & \checkmark \\ \hline
$\alpha>0, M_0>0$ & $\times$ & $\times$ & $\times$ & $\times$ \\ \hline
$\alpha>0, M_0<0$ & \checkmark & $\times$ & $\times$ & \checkmark \\ 
\hline \hline
\end{tabular} 
}
\end{center}
\end{table*}
\end{Prop}
{\it Proof}. 
Equations~(\ref{matter-nonrot}) and (\ref{matter-nonrotating}) give
\begin{align}
\begin{aligned}
&\rho\simeq 2(3+\alpha)M_0r^{\alpha},\qquad\rho+p_1=0,\qquad \rho-p_1\simeq 4(3+\alpha)M_0r^{\alpha},\\
&\rho+p_2=\rho+p_3\simeq -\alpha(3+\alpha)M_0r^{\alpha}-(\alpha+q)(3+\alpha+\beta)M_1r^{\alpha+\beta},\\
&\rho-p_2=\rho-p_3\simeq (3+\alpha)(4+\alpha)M_0r^{\alpha},\\
&\rho+p_1+p_2+p_3\simeq -2(2+\alpha)(3+\alpha)M_0r^{\alpha}
\end{aligned}
\end{align}
around $r=0$.
Then, the proposition follows from Eqs.~(\ref{NEC})--(\ref{SEC}).
\qed
%----------------------- lemma ------------------------------%

Proposition~\ref{Prop:no-go-nonrot} shows that at least one of the standard energy conditions must be violated around the regular center.
In particular, as shown by Zaslavskii~\cite{Zaslavskii:2010qz}, only the SEC is violated in the case with $\alpha=0$, $M_0>0$, and $M_1<0$.
This implies that gravity becomes repulsive around the regular center, which is not a fatal problem as the violations of other energy conditions. 

On the other hand, the spacetime (\ref{metric-app}) is asymptotically flat as $r\to \infty$ if the mass function obeys $M(r)={O}(r^0)$ there.
We present a simple criterion to check the energy conditions in an asymptotically flat region.
%----------------------- lemma ------------------------------%
\begin{Prop}
\label{Prop:no-go-nonrot-infty}
Suppose that $M(r)$ in the spacetime (\ref{metric-app}) is expanded in an asymptotically flat region $r\to\infty$ as 
\begin{align}
M(r)\simeq \frac{M_0}{r^{\alpha}}+\frac{M_1}{r^{\alpha+\beta}}+\frac{M_2}{r^{\alpha+\beta+\gamma}} \label{M-exp-infty}
\end{align}
with $\alpha\ge 0$, $\beta>0$, $\gamma>0$, and $M_0M_1M_2\ne 0$.
Then, the standard energy conditions are respected and violated near $r\to \infty$ as shown in Table~\ref{table:EC-infty}.
\begin{table*}[htb]
\begin{center}
\caption{\label{table:EC-infty} The energy conditions near $r\to \infty$ in the spacetime (\ref{metric-app}) with Eq.~(\ref{M-exp-infty}).}
\scalebox{1.0}{
\begin{tabular}{|c||c|c|c|c|c|}
\hline \hline
& {\rm NEC} & {\rm WEC} & {\rm DEC} & {\rm SEC} \\\hline
$\alpha=0, M_1>0$ & $\times$ & $\times$ & $\times$ & $\times$ \\ \hline
$\alpha=0, \beta> 1, M_1<0$ & \checkmark & \checkmark & $\times$ & \checkmark \\ \hline
$\alpha=0, 0<\beta< 1, M_1<0$ & \checkmark & \checkmark & $\checkmark$ & \checkmark \\ \hline
$\alpha=0, \beta=1, M_1<0, M_2>0$ & \checkmark & \checkmark & \checkmark & \checkmark \\ \hline
$\alpha=0, \beta=1, M_1<0, M_2<0$ & \checkmark & \checkmark & $\times$ & \checkmark \\ \hline
$\alpha>0, M_0>0$ & $\times$& $\times$ & $\times$ & $\times$ \\ \hline
$\alpha>0, M_0<0$ & \checkmark & \checkmark & $\times$ & \checkmark \\ 
\hline \hline
\end{tabular} 
}
\end{center}
\end{table*}
\end{Prop}
{\it Proof}. 
Equations~(\ref{matter-nonrot}) and (\ref{matter-nonrotating}) give
\begin{align}
\label{EC-infty-expand}
\begin{aligned}
&\rho\simeq -\frac{2\alpha M_0}{r^{\alpha+3}}-\frac{2(\alpha+\beta)M_1}{r^{\alpha+\beta+3}},\\
&\rho+p_1=0,\qquad \rho-p_1\simeq -\frac{4\alpha M_0}{r^{\alpha+3}}-\frac{4(\alpha+\beta)M_1}{r^{\alpha+\beta+3}},\\
&\rho+p_2=\rho+p_3\simeq -\frac{\alpha(\alpha+3)M_0}{r^{\alpha+3}}-\frac{(\alpha+\beta)(\alpha+\beta+3)M_1}{r^{\alpha+\beta+3}},\\
&\rho-p_2=\rho-p_3\simeq \frac{\alpha(\alpha-1)M_0}{r^{\alpha+3}}+\frac{(\alpha+\beta)(\alpha+\beta-1)M_1}{r^{\alpha+\beta+3}} \\
&~~~~~~~~~~~~~~~~~~~~~~~~~~~~~~~+\frac{(\alpha+\beta+\gamma)(\alpha+\beta+\gamma-1)M_2}{r^{\alpha+\beta+\gamma+3}},\\
&\rho+p_1+p_2+p_3\simeq -\frac{2\alpha(\alpha+1)M_0}{r^{\alpha+3}}-\frac{2(\alpha+\beta)(\alpha+\beta+1)M_1}{r^{\alpha+\beta+3}}
\end{aligned}
\end{align}
near $r\to \infty$.
Then, the proposition follows from Eqs.~(\ref{NEC})--(\ref{SEC}).
\qed
%----------------------- lemma ------------------------------%

\subsubsection{Rotating case}

We will also study the rotating counterparts of the spherically symmetric black holes described by the metric (\ref{metric-app}).
In particular, we assume that such rotating counterparts are described by the following stationary and axisymmetric G\"urses-G\"ursey metric~\cite{Gurses:1975vu}:
\begin{align}
\label{BL}
\begin{aligned}
\D s^2=&-\biggl(1-\frac{2M(r)r}{\Sigma(r,\theta)}\biggl)\D t^2-\frac{4aM(r)r\sin^2\theta}{\Sigma(r,\theta)}\D t\D\phi \\
&+\frac{\Sigma(r,\theta)}{\Delta(r)}\D r^2+\Sigma(r,\theta)\D\theta^2+\biggl(r^2+a^2+\frac{2a^2M(r)r\sin^2\theta}{\Sigma(r,\theta)}\biggl)\sin^2\theta\D\phi^2, \\
&\Sigma(r,\theta):=r^2+a^2\cos^2\theta,\qquad \Delta(r):=r^2+a^2-2rM(r).
\end{aligned}
\end{align} 
Here $a$ is a constant characterizing the angular momentum of the spacetime and the metric (\ref{BL}) reduces to Eq.~(\ref{metric-app}) for $a=0$.
The metric (\ref{BL}) satisfies $g_{tt}g_{\phi\phi}-g_{t\phi}^2=-\Delta(r)\sin^2\theta$ and $g_{tt}(r,\theta)=0$ determines the location of an ergosphere $r=r_{\rm erg}(\theta)$.
A regular null hypersurface $r=r_{\rm h}$ satisfying $\Delta(r_{\rm h})=0$ is a Killing horizon associated with a Killing vector $\xi^\mu=(1,0,0,a/(r_{\rm h}^2+a^2))$, of which squared norm is 
\begin{align}
\xi_\mu\xi^\mu=\frac{2rM(r)(r^2-r_{\rm h}^2)^2-\Sigma(r,\theta)\{\Delta(r)\Sigma(r,\theta)+(r^2-r_{\rm h}^2)(r^2-r_{\rm h}^2-2\Delta(r))\}}{(r_{\rm h}^2+a^2)^2\Sigma(r,\theta)}.
\end{align}
Similar to the spherically symmetric case, a Killing horizon is referred to as {\it outer} if $\D \Delta/\D r|_{r=r_{\rm h}}>0$, {\it inner} if $\D \Delta/\D r|_{r=r_{\rm h}}<0$, and {\it degenerate} if $\D \Delta/\D r|_{r=r_{\rm h}}=0$.

In the coordinate system (\ref{BL}), a Killing horizon defined by $\Delta(r_{\rm h})=0$ is a coordinate singularity, on which $r_{\rm h}M(r_{\rm h})>0$ holds.
A region with $rM(r)\ge 0$ in the G\"urses-G\"ursey spacetime (\ref{BL}) can be expressed in the following Doran coordinates $({\bar t},r,\theta,\varphi)$~\cite{Doran:1999gb,Visser:2007fj}:
\begin{align}
\D s^2=&-\D {\bar t}^2+\Sigma(r,\theta)\D \theta^2+(r^2+a^2)\sin^2\theta\D\varphi^2\nonumber \\
&+\frac{\Sigma(r,\theta)}{r^2+a^2}\biggl\{\D r+\frac{\sqrt{2M(r)r(r^2+a^2)}}{\Sigma(r,\theta)}(\D {\bar t}-a\sin^2\theta\D\varphi)\biggl\}^2.\label{Doran}
\end{align}
In the coordinates system (\ref{Doran}), a Killing horizon is not a coordinate singularity and $g^{{\bar t}{\bar t}}=-1$ holds.
With $a=0$, the metric (\ref{Doran}) reduces to the spherically symmetric metric (\ref{metric-app}) in the Painlev\'{e}-Gullstrand coordinates.
In the Petrov classification, the spacetime (\ref{Doran}) is of type D.
%----------------------- lemma ------------------------------%
\begin{Prop}
\label{Prop:Petrov-D}
The G\"urses-G\"ursey spacetime (\ref{Doran}) is of Petrov type D.
\end{Prop}
{\it Proof:}
We introduce a complex null tetrad $\{l^\mu,n^\mu,m^\mu,{\bar m}^\mu\}$ as
\begin{align}
&l_\mu\D x^\mu=\frac{1}{\sqrt{2}}\biggl[-\D {\bar t}-\sqrt{\frac{\Sigma}{r^2+a^2}}\biggl\{\D r+\frac{\sqrt{2M(r)r(r^2+a^2)}}{\Sigma}(\D {\bar t}-a\sin^2\theta\D\varphi)\biggl\}\biggl],\\
&n_\mu\D x^\mu=\frac{1}{\sqrt{2}}\biggl[-\D {\bar t}+\sqrt{\frac{\Sigma}{r^2+a^2}}\biggl\{\D r+\frac{\sqrt{2M(r)r(r^2+a^2)}}{\Sigma}(\D {\bar t}-a\sin^2\theta\D\varphi)\biggl\}\biggl],\\
&m_\mu\D x^\mu=\frac{1}{\sqrt{2}}\biggl(\sqrt{\Sigma}\D \theta+i\sqrt{r^2+a^2}\sin\theta\D\varphi\biggl),\\
&{\bar m}_\mu\D x^\mu=\frac{1}{\sqrt{2}}\biggl(\sqrt{\Sigma}\D \theta-i\sqrt{r^2+a^2}\sin\theta\D\varphi\biggl),
\end{align} 
which satisfy $l_\mu n^\mu=-1$, $m_\mu {\bar m}^\mu=1$, and $g_{\mu\nu}=-l_\mu n_\nu-n_\mu l_\nu+m_\mu{\bar m}_\nu+{\bar m}_\mu m_\nu$.
Then, the Weyl scalars are computed to give
\begin{align}
\Psi_0 :=& C_{\alpha\beta\gamma\delta} l^\alpha m^\beta l^\gamma m^\delta \nonumber \\
=&a^2\sin^2\theta\biggl(\frac{ rM''}{4\Sigma^2}-\frac{(2ir+a\cos\theta)M'}{2\Sigma^2(ir-a\cos\theta)}-\frac{3iM}{2\Sigma(ir-a\cos\theta)^3}\biggl)\ , \label{Psi0}\\
\Psi_1 :=& C_{\alpha\beta\gamma\delta} l^\alpha n^\beta l^\gamma m^\delta=\frac{i\sqrt{r^2+a^2}}{a\sin\theta}\Psi_0\ , \\
\Psi_2 :=& C_{\alpha\beta\gamma\delta} l^\alpha m^\beta \bar{m}^\gamma n^\delta =-\frac{2r^2+3a^2-a^2\cos^2\theta}{3a^2\sin^2\theta}\Psi_0\ , \\
\Psi_3 :=& C_{\alpha\beta\gamma\delta} l^\alpha n^\beta \bar{m}^\gamma n^\delta=-\Psi_1\ , \\
\Psi_4 :=& C_{\alpha\beta\gamma\delta} n^\alpha \bar{m}^\beta n^\gamma \bar{m}^\delta=\Psi_0\ ,\label{Psi4}
\end{align} 
with which an invariant algebraic equation $\Psi_4Z^4+4\Psi_3 Z^3+6\Psi_2Z^2+4\Psi_1Z+\Psi_0=0$ admits solutions $Z=Z_1, Z_2, Z_3, Z_4$, where
\begin{align}
\begin{aligned}
Z_1=&i\frac{\sqrt{r^2+a^2}-\sqrt{\Sigma}}{a\sin\theta},\qquad Z_2=i\frac{\sqrt{r^2+a^2}+\sqrt{\Sigma}}{a\sin\theta},\\
Z_3=&i\frac{\sqrt{r^2+a^2}-\sqrt{\Sigma}}{a\sin\theta},\qquad Z_4=i\frac{\sqrt{r^2+a^2}+\sqrt{\Sigma}}{a\sin\theta}.
\end{aligned} 
\end{align} 
Since $Z_1= Z_3\ne Z_2= Z_4$ holds, the spacetime is of Petrov type D. (See Sec.~9 in the textbook~\cite{Chandrasekhar:1985kt}.)
\qed
%----------------------- lemma ------------------------------%

A time orientable spacetime is said to be {\it causal} if there is no closed causal curve~\cite{Hawking:1973uf}.
We will use later the following sufficient condition for the spacetime (\ref{BL}) to be {\it stably causal}, which is stronger than causal and means that no closed causal curve appear even under any small perturbation against the metric.
%----------------------- lemma ------------------------------%
\begin{Prop}
\label{Prop:Causal}
A spacetime described by the G\"urses-G\"ursey metric (\ref{BL}) is stably causal if $rM(r)\ge 0$ holds.
\end{Prop}
{\it Proof}. 
By Proposition 6.4.9 in~\cite{Hawking:1973uf}, a time-orientable spacetime is everywhere stably causal if and only if there is a {\it time function} $T$, which is a differentiable function giving timelike $\nabla_\mu T$.
Since $rM(r)\ge 0$ holds by assumption, we can use the Doran coordinates (\ref{Doran}).
Then, since a vector $U^\mu:=\nabla^\mu {\bar t}$ is everywhere timelike satisfying $U_\mu U^\mu=-1$, the spacetime (\ref{Doran}) is time-orientable by $U^\mu$.
Furthermore, since ${\bar t}$ is a time function, the spacetime is stably causal.
\qed
%----------------------- lemma ------------------------------%

In the Kerr case, where $M(r)$ is constant in the spacetime (\ref{BL}), it is well-known that $(r,\theta)=(0,\pi/2)$ is a ring-like curvature singularity.
In contrast, if the mass function $M(r)$ can be expanded around $r=0$ as $M(r)\simeq M_0 r^{3+\alpha}$ with $M_0\ne 0$ and $\alpha\ge 0$, $(r,\theta)=(0,\pi/2)$ is not a scalar polynomial curvature singularity as shown in the following proposition.

%----------------------- lemma ------------------------------%
\begin{Prop}
\label{Prop:conical}
Suppose that $M(r)$ in the spacetime (\ref{BL}) with $a\ne 0$ is expanded around $r=0$ as $M(r)\simeq M_0 r^{3+\alpha}$ with $M_0\ne 0$ and a non-negative integer $\alpha$.
Then, $(r,\theta)=(0,\pi/2)$ is a ring-like conical singularity and not a scalar polynomial curvature singularity.
The spacetime (\ref{BL}) is locally flat at $r=0$ with $\theta\ne \pi/2$ and can be extended beyond there into the region with negative $r$.
\end{Prop}
{\it Proof:}
By Theorem 1 in~\cite{Torres:2016pgk}, all the second-order curvature invariants are finite at $(r,\theta)=(0,\pi/2)$ in the spacetime (\ref{BL}) with a $C^3$ function $M(r)$ if and only if $M(0)=M'(0)=M''(0)=0$ holds.
Therefore, $(r,\theta)=(0,\pi/2)$ is not a scalar polynomial curvature singularity in the present case.

Under the assumptions, $\lim_{r\to 0}R^{\mu\nu}_{~~~\rho\sigma}=0$ holds with $\theta\ne \pi/2$ and hence the spacetime (\ref{BL}) is locally flat there.
The quantity $\eta:=a\cos\theta/r$ satisfies $\eta\to \infty$ as $r\to 0$ with $\theta\ne \pi/2$.
The value of $\eta$ at $(r,\theta)=(0,\pi/2)$ depends on the path approaching there and can be both finite and infinite.
Because of 
\begin{align}
\lim_{r\to 0}\frac{M(r)r}{\Sigma(r,\theta)}\simeq \frac{M_0r^{2+\alpha}}{1+\eta^2}\to 0
\end{align} 
independent of the value of $\theta$, the spacetime (\ref{BL}) near $r=0$ is given by 
\begin{align}
\D s^2\simeq &-\D t^2+\frac{r^2+a^2\cos^2\theta}{r^2+a^2}\D r^2+(r^2+a^2\cos^2\theta)\D\theta^2+(r^2+a^2)\sin^2\theta\D\phi^2, \label{BL-r=0}
\end{align}
which is Minkowski in the oblate spheroidal coordinates.
Since the metric (\ref{BL}) is of class $C^2$ at $r=0$ with $\theta\ne \pi/2$, the spacetime can be extended beyond $r=0$ into the region with negative $r$ as explained below.

\begin{figure}[htbp]
\begin{center}
%\rotatebox{-90}{
\includegraphics[width=1.0\linewidth]{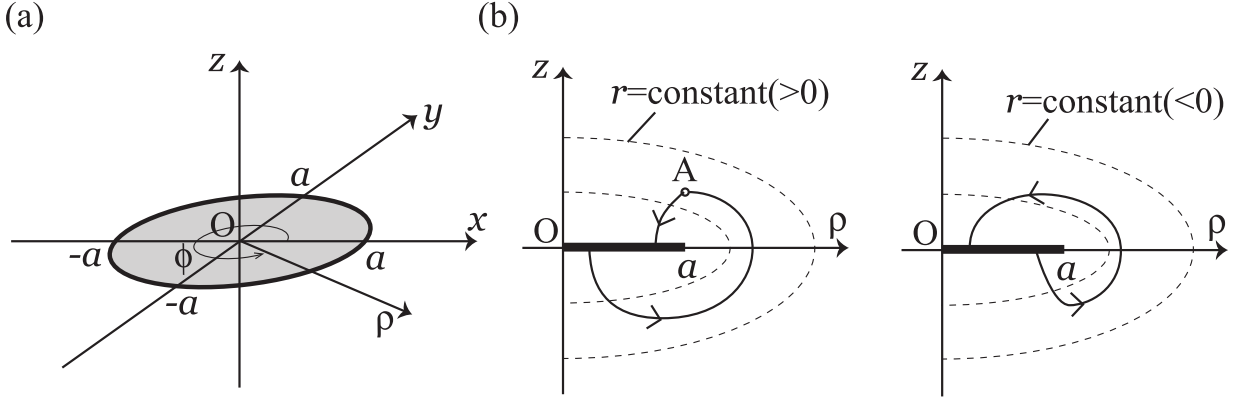}
%\subfigure[]{\includegraphics[width=0.7\linewidth]{Roberts-lambda1.eps}}
%\subfigure[]{\includegraphics[width=0.7\linewidth]{Roberts-lambda2.eps}}
%}
\caption{\label{ConicalSingularity} (a) The Cartesian coordinates $(x,y,z)$ and the cylindrical coordinates $(\rho,\phi,z)$ and (b) the domains of $r>0$ (left) and $r<0$ (right). A closed curve $\gamma$ from a point A wraps around a ring at $(\rho,z)=(a,0)$ by crossing twice a disk defined by $\rho\in[0,a)$ with $z=0$.
}
\end{center}
\end{figure}

The metric (\ref{BL-r=0}) is obtained from the following flat metric in the cylindrical coordinates $(t,\rho,\phi,z)$
\begin{align}
\D s^2=-\D t^2+\D\rho^2+\rho^2\D\phi^2+\D z^2\label{cylinder}
\end{align}
by the coordinate transformations
\begin{align}
\rho=\sqrt{r^2+a^2}\sin\theta,\qquad z=r\cos\theta
\end{align}
which satisfies
\begin{align}
\frac{\rho^2}{r^2+a^2}+\frac{z^2}{r^2}=1.\label{spheroid}
\end{align}
Now we follow the same argument in~\cite{Gibbons:2017djb}.
Equation~(\ref{spheroid}) shows that $r=$constant represents a spheroid ($0\le \rho<\infty$ and $-\infty<z<\infty$) as shown in Fig.~\ref{ConicalSingularity}.
In the limit $r\to 0$, this spheroid reduces to a segment $\rho\in[0,a]$ with $z=0$, while $z=0$ outside this segment corresponds to $\theta=\pi/2$ with $r^2=\rho^2-a^2$. (See Fig.~\ref{ConicalSingularity}.)
Equation~(\ref{spheroid}) is solved to give 
\begin{align}
r^2=&\frac12\left\{(\rho^2+z^2-a^2)+\sqrt{(\rho^2+z^2-a^2)^2+4a^2z^2}\right\}(\ge 0),\label{r2}
\end{align}
of which limit to $z\to 0$ depends on the regions as
\begin{equation}
\lim_{z\to 0}r^2\simeq
\begin{cases}
\displaystyle\frac{a^2}{a^2-\rho^2}z^2+{O}(z^4) & (\mbox{if~}\rho^2< a^2),\\
\displaystyle(\rho^2-a^2)+{O}(z^2) & (\mbox{if~}\rho^2> a^2).
\end{cases}
\end{equation}
This shows
\begin{equation}
\lim_{z\to 0}\cos\theta= \lim_{z\to 0}\frac{z}{r}\simeq 
\begin{cases}
\displaystyle\sqrt{1-\frac{\rho^2}{a^2}}\frac{z}{|z|} & (\mbox{if~}\rho^2< a^2),\\
\displaystyle 0 & (\mbox{if~}\rho^2> a^2)
\end{cases}
\end{equation}
in the domain of $r\ge 0$ and 
\begin{equation}
\lim_{z\to 0}\cos\theta= \lim_{z\to 0}\frac{z}{r}\simeq 
\begin{cases}
\displaystyle -\sqrt{1-\frac{\rho^2}{a^2}}\frac{z}{|z|} & (\mbox{if~}\rho^2< a^2),\\
\displaystyle 0 & (\mbox{if~}\rho^2> a^2).
\end{cases}
\end{equation}
in the domain of $r\le 0$, where $z/|z|=\pm 1$ as $z\to \pm 0$.
Hence, in order to remove the discontinuity at $z=0$, the extension of the spacetime beyond the ``disk'' described by $\rho\in[0,a)$ with $z=0$ requires attachment to a spacetime region with negative $r$.
As a consequence, in order to draw a closed curve wrapping around a ``ring'' described by $(\rho,z)=(a,0)$, the curve has to intersect the disk at least twice. (See Fig.~\ref{ConicalSingularity}.)
Therefore, this ring $(\rho,z)=(a,0)$, or equivalently $(r,\theta)=(0,\pi/2)$, is a conical singularity~\cite{Gibbons:2017djb}.

As a complement, we show that $(r,\theta)=(0,\pi/2)$ is a conical singularity in a different manner~\cite{Gibbons:2017djb}.
With the coordinates $x_1:=r/a$ and $x_2:=\cos\theta$, the metric (\ref{BL-r=0}) reduces for small $x_1$ and $x_2$ (near the ring $(r,\theta)=(0,\pi/2)$) to
\begin{align}
\D s^2\simeq &-\D t^2+a^2(x_1^2+x_2^2)(\D x_1^2+\D x_2^2)+a^2\D\phi^2 \nonumber \\
=&-\D t^2+\D x^2+x^2\D{\bar\theta}^2+a^2\D\phi^2,\label{BL-r=0-x12}
\end{align}
where coordinates $x$ and ${\bar\theta}$ are defined by $x_1=\sqrt{2x/a}\cos({\bar\theta}/2)$ and $x_2=\sqrt{2x/a}\cos({\bar\theta}/2)$ satisfying $(x_1+ix_2)^2=(2/a)xe^{i{\bar\theta}}$.
Since the domains of $x$ and ${\bar\theta}$ are $x\in[0,\infty)$ and ${\bar\theta}\in [0,4\pi]$, the metric (\ref{BL-r=0-x12}) contains a conical singularity at $x= 0$ stretching along the $\phi$-direction~\cite{Gibbons:2017djb}.
\qed
%----------------------- lemma ------------------------------%

%------------<fig>---------------------------
%--------------<fig>-----------------------

Although Proposition~\ref{Prop:conical} shows that $(r,\theta)=(0,\pi/2)$ is not a scalar polynomial curvature singularity, it can still be a parallelly p.p. curvature singularity~\cite{Hawking:1973uf}.
In fact, while $\lim_{r\to 0}R^{\mu\nu}_{~~~\rho\sigma}=2M_0(\delta^\mu_{\rho}\delta^\nu_{\sigma}-\delta^\mu_{\sigma}\delta^\nu_{\rho})$ holds on the equatorial plane $\theta=\pi/2$ for $\alpha=0$, some components of the Riemann tensor $R^{\mu\nu}_{~~~\rho\sigma}$ blow up as $r\to 0$ on the equatorial plane for $\alpha>0$, which might be a sign of p.p. curvature singularity.
If $(r,\theta)=(0,\pi/2)$ is not a p.p. curvature singularity, rotating non-singular black holes described by the G\"urses-G\"ursey metric (\ref{BL}) under the assumptions in Proposition~\ref{Prop:conical} do not satisfy the criterion C1 but satisfy W-C1 in Section~\ref{sec:intro}.
We leave this problem for future investigations.

In the following section, we will also check the energy conditions for ${\tilde T}_{\mu\nu}$ in the spacetime (\ref{BL}).
The rotating spacetime (\ref{BL}) may be written as
\begin{align}
\label{BL2}
\begin{aligned}
\D s^2=&-\frac{\Delta(r)}{\Sigma(r,\theta)}\left(\D t-a\sin^2\theta\D\phi\right)^2+\frac{\Sigma(r,\theta)}{\Delta(r)}\D r^2 \\
&+\Sigma(r,\theta)\D\theta^2+\frac{\sin^2\theta}{\Sigma(r,\theta)}\left\{a\D t-(r^2+a^2)\D\phi\right\}^2,
\end{aligned}
\end{align} 
which shows the following natural set of orthonormal basis one-forms:
\begin{align}
\label{basis-rotating}
\begin{aligned}
&E^{(0)}_\mu\D x^\mu = \left\{
\begin{array}{ll}
-\sqrt{\Delta/\Sigma}(\D t-a \sin^2\theta\D\phi) & (\mbox{if}~\Delta(r)>0)\\
-\sqrt{-\Sigma/\Delta}\D r & (\mbox{if}~\Delta(r)<0)
\end{array}
\right.,\\
&E^{(1)}_\mu\D x^\mu = \left\{
\begin{array}{ll}
-\sqrt{\Sigma/\Delta}\D r & (\mbox{if}~\Delta(r)>0)\\
-\sqrt{-\Delta/\Sigma}(\D t-a \sin^2\theta\D\phi) & (\mbox{if}~\Delta(r)<0)
\end{array}
\right.,\\
&E^{(2)}_\mu\D x^\mu=\sqrt{\Sigma}\D \theta,\qquad E^{(3)}_\mu\D x^\mu=\frac{\sin\theta}{\sqrt{\Sigma}}\left\{-a\D t+(r^2+a^2)\D\phi\right\},
\end{aligned}
\end{align}
which satisfy $\eta^{(a)(b)}=g^{\mu\nu}{E}_\mu^{(a)} {E}_\nu^{(b)}=\mbox{diag}(-1,1,1,1)$.
Then, as shown in~\cite{Gurses:1975vu,Burinskii:2001bq}, ${\tilde T}^{(a)(b)}={\tilde T}^{\mu\nu}{E}_\mu^{(a)} {E}_\nu^{(b)}$ is given in the type-I form~(\ref{T-typeI}) with
\begin{align}
\rho=&-p_1=\frac{2r^2M'}{\Sigma^2},\qquad p_2=p_3=-\frac{rM''\Sigma+2M'a^2\cos^2\theta}{\Sigma^2} \label{matter-rot}
\end{align}
independent of the sign of $\Delta$, which show
\begin{align}
\label{matter-rotating}
\begin{aligned}
&\rho+p_1=0,\qquad \rho-p_1=\frac{4 r^2M'}{\Sigma^2},\\
&\rho+p_2=\rho+p_3=\frac{2M'(r^2-a^2\cos^2\theta)-rM''\Sigma}{\Sigma^2},\\
&\rho-p_2=\rho-p_3=\frac{2M'+rM''}{\Sigma},\\
&\rho+p_1+p_2+p_3=-\frac{2(rM''\Sigma+2M'a^2\cos^2\theta)}{\Sigma^2}.
\end{aligned}
\end{align}

Since a Killing horizon defined by $\Delta(r_{\rm h})=0$ is a coordinate singularity in the coordinate system (\ref{BL}), the basis one-forms (\ref{basis-rotating}) are not well-defined there.
Nevertheless, Eqs.~(\ref{matter-rot}) and (\ref{matter-rotating}) are still valid on a Killing horizon $r=r_{\rm h}$ by the following lemma.
%----------------------- lemma ------------------------------%
\begin{lm}
\label{lm:horizon-matter-rotating}
Equations~(\ref{matter-rot}) and (\ref{matter-rotating}) are valid on a Killing horizon $\Delta(r_{\rm h})=0$ in the G\"urses-G\"ursey spacetime (\ref{BL}).
\end{lm}
{\it Proof}. 
Since $r_{\rm h}M(r_{\rm h})\ge 0$ holds on a Killing horizon, we can use the Doran coordinates (\ref{Doran}).
A natural set of the basis one-forms in the spacetime (\ref{Doran}) consists of
\begin{align}
E_\mu^{(0)}\D x^\mu=&-\D {\bar t},\\
E_\mu^{(1)}\D x^\mu=&\sqrt{\frac{\Sigma(r,\theta)}{r^2+a^2}}\biggl\{\D r+\frac{\sqrt{2M(r)r(r^2+a^2)}}{\Sigma(r,\theta)}(\D {\bar t}-a\sin^2\theta\D\varphi)\biggl\},\\
E_\mu^{(2)}\D x^\mu=&\Sigma(r,\theta)^{1/2}\D\theta,\\
E_\mu^{(3)}\D x^\mu=&\sqrt{r^2+a^2}\sin\theta\D\varphi,
\end{align}
with which ${\tilde T}^{(a)(b)}$ admits off-diagonal components ${\tilde T}^{(0)(3)}(={\tilde T}^{(3)(0)})$.
In order to remove them, we introduce a new set of the basis one-forms $\{{\bar E}_\mu^{(0)},{E}_\mu^{(1)},{E}_\mu^{(2)},{\bar E}_\mu^{(3)}\}$ obtained by a local Lorentz boost on the plane spanned by $E_\mu^{(0)}$ and $E_\mu^{(3)}$ such that 
\begin{align}
{\bar E}_\mu^{(0)}=&\cosh\Theta E_\mu^{(0)}-\sinh\Theta E_\mu^{(3)},\\
{\bar E}_\mu^{(3)}=&-\sinh\Theta E_\mu^{(0)}+\cosh\Theta E_\mu^{(3)}
\end{align}
with
\begin{align}
\cosh\Theta=\sqrt{\frac{r^2+a^2}{\Sigma(r,\theta)}},\qquad \sinh\Theta=-\frac{a\sin\theta}{\sqrt{\Sigma(r,\theta)}}.
\end{align}
With the new set of basis one-forms, the non-zero components of ${\tilde T}^{(a)(b)}$ are
\begin{align}
{\tilde T}^{(0)(0)}=&\frac{2r^2M'}{\Sigma^2},\qquad {\tilde T}^{(1)(1)}=-\frac{2r^2M'}{\Sigma^2},\\
{\tilde T}^{(2)(2)}=&{\tilde T}^{(3)(3)}=-\frac{rM''\Sigma +2a^2 M'\cos^2\theta}{{\Sigma^2}}.
\end{align}
This is the type-I form~(\ref{T-typeI}) with Eq.~(\ref{matter-rot}).
\qed
%----------------------- lemma ------------------------------%

By Proposition~\ref{Prop:center-regularity}, $r=0$ in the spherically symmetric spacetime (\ref{metric-app}) is regular if the mass function behaves around there as $M(r)\simeq M_0 r^{3+\alpha}$ with $M_0\ne 0$ and $\alpha\ge 0$.
With such a mass function, a no-go result is available in the rotating case.
The following proposition was proven with a focus on the WEC in~\cite{Neves:2014aba} for $\alpha=0$ and in~\cite{Torres:2016pgk} for general $\alpha(\ge 0)$.
%----------------------- lemma ------------------------------%
\begin{Prop}
\label{Prop:no-go-rot}
Suppose that $M(r)$ in the spacetime (\ref{BL}) with $a\ne 0$ is expanded around $r=0$ as $M(r)\simeq M_0 r^{3+\alpha}$ with $M_0\ne 0$ and a non-negative integer $\alpha$.
Then, the standard energy conditions are respected and violated near $r=0$ with $\theta\ne \pi/2$ as shown in Table~\ref{table:EC-r=0-rot}.
\begin{table*}[htb]
\begin{center}
\caption{\label{table:EC-r=0-rot} The energy conditions near $r=0$ in the spacetime (\ref{BL}) with $M(r)\simeq M_0 r^{3+\alpha}$.}
\scalebox{1.0}{
\begin{tabular}{|c||c|c|c|c|c|}
\hline \hline
& {\rm NEC} & {\rm WEC} & {\rm DEC} & {\rm SEC} \\\hline
$M_0>0$ & $\times$ & $\times$ & $\times$ & $\times$ \\ \hline
$M_0<0$ & \checkmark & $\times$ & $\times$ & \checkmark \\ 
\hline \hline
\end{tabular} 
}
\end{center}
\end{table*}
\end{Prop}
{\it Proof}. 
Equations~(\ref{matter-rot}) and (\ref{matter-rotating}) with $\theta\ne \pi/2$ give
\begin{align}
\begin{aligned}
&\rho\simeq \frac{2(3+\alpha)M_0}{a^4\cos^4\theta}r^{4+\alpha},\qquad \rho+p_1=0,\qquad \rho-p_1\simeq \frac{4(3+\alpha)M_0}{a^4\cos^4\theta}r^{4+\alpha},\\
&\rho+p_2=\rho+p_3\simeq -\frac{(\alpha+3)(\alpha+4)M_0}{a^2\cos^2\theta}r^{2+\alpha},\\
&\rho-p_2=\rho-p_3\simeq \frac{(\alpha+3)(\alpha+4)M_0}{a^2\cos^2 \theta}r^{2+\alpha},\\
&\rho+p_1+p_2+p_3\simeq -\frac{2(\alpha+3)(\alpha+4)M_0}{a^2\cos^2 \theta}r^{2+\alpha}
\end{aligned}
\end{align}
around $r=0$.
Then, the proposition in the case of $\theta\ne \pi/2$ follows from Eqs.~(\ref{NEC})--(\ref{SEC}).
\qed
%----------------------- lemma ------------------------------%

\noindent
With $\theta=\pi/2$, Eqs.~(\ref{matter-rot}) and (\ref{matter-rotating}) reduce to Eqs.~(\ref{matter-nonrot}) and (\ref{matter-nonrotating}), respectively, and therefore, the standard energy conditions on the equatorial plane $\theta=\pi/2$ in the rotating spacetime (\ref{BL}) are identical to those in the spherically symmetric spacetime (\ref{metric-app}).
Thus, the energy conditions near $r=0$ on the equatorial plane $\theta=\pi/2$ are clarified according to Proposition~\ref{Prop:no-go-nonrot}.

\subsection{A brief history of research}
\label{sec:history}

In 1968, at the 5th International Conference on Gravity and the Theory of Relativity (GR5) held at Tbilisi, Bardeen presented the first spherically symmetric model of a non-singular black hole~\cite{Bardeen1968}.
This Bardeen black hole is of the regular-center type described by the metric (\ref{metric-app}) and has been a prototype of all the subsequent models.
Here we briefly summarize the research history of this type of non-singular black holes and their rotating counterparts described by the metric (\ref{BL}).
We recommend a review paper~\cite{Ansoldi:2008jw} to the readers, which focuses on the results until 2008.
The first section of the paper~\cite{Lemos:2011dq} also provides a nice review of the research history until 2011.

Perhaps without recognizing Bardeen's work, Dymnikova proposed her first model of a non-singular black hole in 1992~\cite{Dymnikova:1992ux}\footnote{Bardeen's paper is not cited in the reference of Dymnikova's paper~\cite{Dymnikova:1992ux}.}, inspired by the work by Poisson and Israel~\cite{Poisson:1988wc} trying to replace the central region of the Schwarzschild black hole by the de~Sitter geometry.
This solution was further studied by herself in~\cite{Dymnikova:1996ijmpd,Dymnikova:1999cz,Dymnikova:2001fb}, of which results were summarized in~\cite{Dymnikova:2003vt}. 
In these papers, Dymnikova already recognized that the DEC can be respected everywhere in this type of spherically symmetric non-singular black holes.
In~\cite{Mars:1996khm}, a different group showed that the WEC can be respected everywhere and presented two explicit models.
Note that Dymnikova's non-singular black-hole spacetime in~\cite{Dymnikova:1992ux}, which is different from the one in~\cite{Dymnikova:2004zc}, has been derived in an iterative renormalization group semi-classical approach~\cite{Platania:2019kyx}.
The energy-density profile of Dymnikova's spacetime in~\cite{Dymnikova:1992ux} has been used to construct a different non-singular black-hole spacetime in~\cite{Mbonye:2005im}, of which thermodynamical and dynamical stabilities were investigated in~\cite{Perez:2014oea}.

In~\cite{Elizalde:2002yz}, the standard energy conditions were studied for several spherically symmetric non-singular black-hole spacetimes with a type-I matter field (\ref{T-typeI}) obeying $p_1=-\rho$ and $p_2=p_3$ and violation of the SEC around a regular center was reported.
A more general result about violation of the SEC was given by Zaslavskii~\cite{Zaslavskii:2010qz}.
In fact, by a contraposition of Theorem 1 in~\cite{Mars:1996gd}, the SEC must be violated somewhere in a spherically symmetric non-singular black-hole spacetime with a regular center.
In~\cite{Balart:2014jia,Balart:2014cga}, more general spherically symmetric models satisfying the WEC were presented, which include models satisfying the DEC.
Further efforts were made along this direction in~\cite{Rodrigues:2017yry}.

In 2006, Hayward proposed a spherically symmetric model of a non-singular black hole in order to provide a possible resolution to the black-hole information-loss conundrum~\cite{Hayward2006}.
This Hayward spacetime reduces to de~Sitter in the limit where the ADM mass parameter is infinitely large and thus satisfies the limiting curvature condition~\cite{Frolov:2016pav,Chamseddine:2016ktu}.
Such non-singular black holes satisfying the limiting curvature condition were investigated in~\cite{Frolov:2016pav,Chamseddine:2016ktu,Chamseddine:2019pux} and also in effective two-dimensional gravity~\cite{Trodden:1993dm,Kunstatter:2015vxa,Frolov:2021kcv}.

\noindent
{\bf Nonlinear electromagnetic field}

Exact solutions representing a non-singular black hole with a regular center have been frequently obtained in general relativity with a class of nonlinear electromagnetic fields, of which action is given by 
\begin{align}
S=\int\D^4 x\sqrt{-g}\biggl(\frac{1}{2}R-\beta {\cal L}(X)\biggl),\label{action-FW}
\end{align} 
where $X:=\alpha F_{\rho\sigma}F^{\rho\sigma}$ and $\beta$ is a coupling constant.
Here we have introduced another constant $\alpha$ to make $X$ dimensionless.
(See Appendix~\ref{App:NEF} for a general class of spherically symmetric solutions in this system in arbitrary $n(\ge 4)$ dimensions.)

In 1998, Ay\'on-Beato and Garc\'{\i}a constructed an electrically charged non-singular black-hole solution with spherical symmetry by solving the field equations not in the system (\ref{action-FW}) but its dual system~\cite{AyonBeato:1998ub} obtained by a Legendre transformation~\cite{Salazar:1987ap}.
The nonlinear electromagnetic field in this solution satisfies the WEC everywhere as well as a proper weak-field limit ${\cal L}(X)\simeq X$ as $X\to 0$ to be the standard Maxwell field.
In their subsequent papers, different electric solutions were obtained again in the dual system~\cite{AyonBeato:1999ec,AyonBeato:1999rg,AyonBeato:2004ih}.
(See also a recent paper~\cite{Cai:2021ele}.)
In 2004, Dymnikova obtained another electric solution also in the dual system, which respects the DEC everywhere~\cite{Dymnikova:2004zc}.

These solutions brought an apparent contradiction to the results by Bronnikov in~\cite{Bronnikov:1979ex,Bronnikov:2000vy} asserting that any system (\ref{action-FW}) admitting a proper Maxwell weak-field limit does not allow electrically charged spherically symmetric static solutions with a regular center.
This apparent paradox was explained in~\cite{Bronnikov:2000vy} that there is no one-to-one mapping from the electric solutions in the dual system obtained in~\cite{AyonBeato:1999ec,AyonBeato:1999rg,AyonBeato:2004ih} to solutions in the original system (\ref{action-FW}).
In other words, in order to describe an electric solution defined in the domain $r\in[0,\infty)$ in a dual system, one needs multiple different Lagrangian functions ${\cal L}(X)$ in the action (\ref{action-FW}) for different ranges of $r$. 

In contrast, magnetic solutions do not suffer from this problem.
In 2000, Ay\'on-Beato and Garc\'{\i}a showed that the Bardeen spacetime can be an exact magnetic solution in the original system (\ref{action-FW})~\cite{AyonBeato:2000zs}.
In 2016, Fan and Wang constructed a wide class of magnetic solutions with spherical symmetry in the system (\ref{action-FW})~\cite{Fan:2016hvf}.
Some solutions in~\cite{Fan:2016hvf} satisfy the DEC everywhere and admit a proper Maxwell weak-field limit.

It should be noted that there is Birkhoff's theorem in the system (\ref{action-FW}) and the general spherically symmetric solution is obtained in the form of Eq.~(\ref{metric-app}) with a mass function $M(r)$ determined by the form of ${\cal L}(X)$. (See Eq.~(\ref{dyo-sol1}) in Appendix~\ref{App:NEF}.)
In this general solution, the mass function $M(r)$ contains an integration constant ${\bar M}$ given from the vacuum sector and a fine-tuning ${\bar M}=0$ is required to generate a non-singular black hole, as pointed out in~\cite{Chinaglia:2017uqd}.
Without a fine-tuning ${\bar M}=0$, there appears a curvature singularity at the center.
Therefore, a non-singular black hole is {\it not} a generic configuration and the criterion C5 in Section~\ref{sec:intro} is not fulfilled in the system (\ref{action-FW}).
This property has been studied in a more general framework in~\cite{Chinaglia:2018gvf}.

In 2003, it was shown that the exterior regions of the non-singular black holes obtained in~\cite{AyonBeato:1998ub,AyonBeato:1999ec,AyonBeato:1999rg,Bronnikov:2000vy,AyonBeato:2000zs} are dynamically stable against gravitational and electromagnetic non-spherical linear perturbations~\cite{Moreno:2002gg}.
In~\cite{Wu:2018xza}, many different spherically symmetric non-singular black holes with a regular center are found to be stable.
In~\cite{Nomura:2020tpc}, sufficient conditions for stability were derived for magnetic black holes in the system with a more general nonlinear electromagnetic field.

\noindent
{\bf Non-Commutative-Geometry inspired models}

In 2006, a spherically symmetric non-singular black hole with a regular center was constructed inspired by noncommutative geometry~\cite{Nicolini:2005vd}, which was later generalized into the electrically charged case~\cite{Ansoldi:2006vg}. 
In these papers, the authors first introduced a modified energy-momentum tensor and then solved the Einstein equations.
As a result, the solutions are described by the metric~(\ref{metric-app}) with certain forms of $M(r)$.
In these Non-Commutative-Geometry inspired models, as a system (\ref{action-FW}) with a nonlinear electromagnetic field, a fine-tuning of the integration constant is required to remove the singularity at the center, so that the criterion C5 in Section~\ref{sec:intro} is not fulfilled.
In~\cite{Smailagic:2010nv}, a rotating counterpart of the solution in~\cite{Nicolini:2005vd} was derived by the Newman-Janis transformation, which is described by the G\"urses-G\"ursey metric (\ref{BL}).

\noindent
{\bf Rotating counterparts}

Back in 1975, G\"urses and G\"ursey established a basis how to construct a rotating counterpart of a spherically symmetric non-singular black hole with a regular center~\cite{Gurses:1975vu}.
They derived a stationary and axisymmetric metric in the Boyer-Lindquist coordinates by the Newman-Janis complex transformation from a metric of the Kerr-Schild class~\cite{Newman:1965tw,Newman:1965my,Drake:1998gf}.
The resulting G\"urses-G\"ursey metric~(\ref{BL}) reduces to the Kerr spacetime when the mass function $M(r)$ is constant~\cite{Gurses:1975vu}.
G\"urses and G\"ursey showed that the corresponding energy-momentum tensor is of the Hawking-Ellis type I given by Eq.~(\ref{T-typeI}) with $p_1=-\rho$ and $p_2=p_3$~\cite{Gurses:1975vu}.
Since then, for its simplicity, the G\"urses-G\"ursey metric~(\ref{BL}) has been widely used with different forms of $M(r)$ to construct rotating counterparts of spherically symmetric non-singular black holes.

In 2002, violation of the WEC was reported in rotating non-singular black holes described by the G\"urses-G\"ursey metric in the Kerr-Schild form~\cite{Burinskii:2001bq}.
In 2013, Bambi and Modesto introduced a generalized G\"urses-G\"ursey metric based on the Newman-Janis transformation, where the mass function $M(r,\theta)$ depends also on an angular coordinate $\theta$~\cite{Bambi:2013ufa}.
They showed that the WEC is violated in the central regions of the rotating counterparts of the Bardeen and Hayward black holes~\cite{Bambi:2013ufa}.
After being confirmed for a more general mass function $M(r)$ with a cosmological constant in~\cite{Neves:2014aba}, violation of the WEC in the central region was shown to be generic in 2016~\cite{Torres:2016pgk}.
Rotating counterparts of the Ay\'on-Beato-Garc\'{\i}a black holes~\cite{AyonBeato:1998ub,AyonBeato:1999ec,AyonBeato:1999rg} 
and the Fan-Wang black hole~\cite{Fan:2016hvf} were investigated in~\cite{Toshmatov:2014nya,Azreg-Ainou:2014pra} and~\cite{Toshmatov:2017zpr}, respectively.
Rotating non-singular black holes described by the G\"urses-G\"ursey metric~(\ref{BL}) with even more different mass functions were investigated in~\cite{Ghosh:2014pba,Amir:2016cen,Ghosh:2020ece}.
A more general stationary and axisymmetric metric than the G\"urses-G\"ursey metric has been discussed by Azreg-A\"\i{}nou in~\cite{Azreg-Ainou:2014pra,Azreg-Ainou:2014nra,Azreg-Ainou:2014aqa}.

%======================================%
%<<<<<<<<<<<< SECTION 1 >>>>>>>>>>>>>>%
%======================================%
\section{Quest for realistic non-singular black-holes}
\label{sec:RC}

In this section, we seek non-singular black holes of the regular-center type which satisfy all the geometric criteria C1--C5 in Sec.~\ref{sec:intro}.
In particular, we study the Bardeen spacetime~\cite{Bardeen1968}, Hayward spacetime~\cite{Hayward2006}, Dymnikova spacetime~\cite{Dymnikova:2004zc}, and Fan-Wang spacetime~\cite{Fan:2016hvf} and their rotating counterparts.
These spacetimes and their rotating counterparts are described by the metric (\ref{metric-app}) and the G\"urses-G\"ursey metric (\ref{BL}), respectively, with the following forms of the mass function $M(r)$:
\begin{align}
\mbox{Bardeen}:~&M(r)=\frac{mr^3}{(r^{2}+l^2)^{3/2}},\label{m-B}\\
\mbox{Hayward}:~&M(r)=\frac{mr^3}{r^{3}+2ml^2}, \label{m-Hayward}\\
\mbox{Dymnikova}:~&M(r)=\frac{2m}{\pi}\biggl\{\arctan\biggl(\frac{r}{l}\biggl)-\frac{lr}{r^2+l^2}\biggl\}, \label{m-D}\\
\mbox{Fan-Wang}:~&M(r)=\frac{mr^3}{(r+l)^3}.\label{m-FW}
\end{align}
These spacetimes are characterized by two parameters $m$ and $l$.
All the mass functions obey $\lim_{r\to\infty}M(r)=m$, so that all these spacetimes are asymptotically flat as $r\to \infty$ and $m$ is the ADM mass.
The spacetimes reduce to Minkowski for $m=0$ and to Schwarzschild for $l=0$ because of $\lim_{l\to 0}M(r)=m$.
We assume $l>0$ throughout this paper.

Since the curvature invariants of the Bardeen, Dymnikova, and Fan-Wang spacetimes blow up as $m\to \infty$, the criterion C5 in Sec.~\ref{sec:intro} is not respected.
In contrast, as pointed out in~\cite{Frolov:2016pav}, the curvature invariants of the Hayward spacetime (with $l\ne 0$) are finite for any values of $m$ and $l$, so that the criterion C5 is respected.

\subsection{Non-rotating case}

Let us first study the standard energy conditions for the effective energy-momentum tensor ${\tilde T}_{\mu\nu}(:=G_{\mu\nu})$ near the regular center $r=0$ and an asymptotically flat region $r\to \infty$.
The mass functions are expanded near $r=0$ as
\begin{align}
\mbox{Bardeen}:~&M(r)\simeq \frac{m}{l^3}r^3-\frac{3m}{2l^5}r^5+O(r^{7}),\label{m-B-r=0}\\
\mbox{Hayward}:~&M(r)\simeq \frac{1}{2l^2}r^3-\frac{1}{4ml^4}r^6+O(r^{9}), \label{m-Hayward-r=0}\\
\mbox{Dymnikova}:~&M(r)\simeq \frac{4m}{3\pi l^3}r^3-\frac{8m}{5\pi l^5}r^5+O(r^{7}), \label{m-D-r=0}\\
\mbox{Fan-Wang}:~&M(r)\simeq \frac{m}{l^3}r^3-\frac{3m}{l^4}r^4+O(r^{5}).\label{m-FW-r=0}
\end{align}
Consistent with the result in~\cite{Zaslavskii:2010qz}, the SEC is violated around the regular center $r=0$ in these spacetimes with $m>0$ by Proposition~\ref{Prop:no-go-nonrot}.
On the other hand, the mass functions are expanded near $r=\infty$ as
\begin{align}
\mbox{Bardeen}:~&M(r)\simeq m-\frac{3ml^2}{2r^2}+O(r^{-4}),\label{m-B-r=infty}\\
\mbox{Hayward}:~&M(r)\simeq m-\frac{2m^2l^2}{r^3}+O(r^{-6}), \label{m-Hayward-r=infty}\\
\mbox{Dymnikova}:~&M(r)\simeq m-\frac{4ml}{\pi r}+\frac{8ml^3}{3\pi r^3}+O(r^{-5}), \label{m-D-r=infty}\\
\mbox{Fan-Wang}:~&M(r)\simeq m-\frac{3ml}{r}+\frac{6ml^2}{r^2}+O(r^{-3}).\label{m-FW-r=infty}
\end{align}
By Proposition~\ref{Prop:no-go-nonrot-infty}, the Dymnikova and Fan-Wang spacetimes with $m>0$ satisfy all the standard energy conditions in an asymptotically flat region $r\to \infty$.
In contrast, the Bardeen and Hayward spacetimes with $m>0$ violate the DEC there.

In order to clarify the global structure of the spacetime (\ref{metric-app}), we will used the following lemma, which was introduced in~\cite{Torii:2005xu}.
%----------------------- lemma ------------------------------%
\begin{lm}
\label{lm:horizon-nonrotating}
Identify the metric function $f(r)$ in the spacetime (\ref{metric-app}) as a function ${\bar f}(r,m)$ of $r$ and $m$ such that ${\bar f}(r,m)\equiv f(r)$.
Define a function $m=m_{\rm h}(r_{\rm h})$ as a solution of ${\bar f}(r_{\rm h},m)=0$, where $r=r_{\rm h}(>0)$ is the radius of a Killing horizon, and suppose that $\partial{\bar f}/\partial m|_{r=r_{\rm h}}$ is non-zero and finite.
Then, in the case of $\partial{\bar f}/\partial m|_{r=r_{\rm h}}<(>)0$, a Killing horizon is outer if $\D m_{\rm h}/\D r_{\rm h}>(<)0$, inner if $\D m_{\rm h}/\D r_{\rm h}<(>)0$, and degenerate if $\D m_{\rm h}/\D r_{\rm h}=0$.
\end{lm}
{\it Proof}. 
The total derivative of the constraint ${\bar f}(r_{\rm h},m_{\rm h}(r_{\rm h}))=0$ gives
\begin{align}
\frac{\D f}{\D r}\biggl|_{r=r_{\rm h}}=-\frac{\partial{\bar f}}{\partial m}\biggl|_{r=r_{\rm h}}\frac{\D m_{\rm h}}{\D r_{\rm h}},\label{diagram1}
\end{align}
from which the lemma follows.
\qed
%----------------------- lemma ------------------------------%

The mass functions (\ref{m-B})--(\ref{m-FW}) give
\begin{align}
\mbox{Bardeen}:~&\frac{\partial{\bar f}}{\partial m}=-\frac{2r^2}{(r^{2}+l^2)^{3/2}},\label{m-B2}\\
\mbox{Hayward}:~&\frac{\partial{\bar f}}{\partial m}=-\frac{2r^5}{(r^3+2ml^2)^{2}}, \label{m-Hayward2}\\
\mbox{Dymnikova}:~&\frac{\partial{\bar f}}{\partial m}=-\frac{4\{(r^2+l^2)\arctan(r/l)-lr\}}{\pi r(r^{2}+l^2)}, \label{m-D2}\\
\mbox{Fan-Wang}:~&\frac{\partial{\bar f}}{\partial m}=-\frac{2r^2}{(r+l)^{3}},\label{m-FW2}
\end{align}
which satisfy $\partial{\bar f}/\partial m<0$ in the domain $r\in(0,\infty)$.
Hence, by Lemma~\ref{lm:horizon-nonrotating}, a Killing horizon is outer if $\D m_{\rm h}/\D r_{\rm h}>0$, inner if $\D m_{\rm h}/\D r_{\rm h}<0$, and degenerate if $\D m_{\rm h}/\D r_{\rm h}=0$.
Now let us study the four spacetimes separately.

\subsubsection{Bardeen spacetime}

The metric function $f(r)$ for the Bardeen spacetime~\cite{Bardeen1968} is 
\begin{align}
f(r)=1-\frac{2mr^2}{(r^{2}+l^2)^{3/2}}. \label{Bardeen-BH-4-ap}
\end{align}
The metric is invariant for $r\to -r$ and the spacetime with $l\ne 0$ is analytic everywhere in the domain $r\in[0,\infty)$.
The metric function $f(r)$ behaves around $r=0$ as $f(r)\simeq 1-2mr^2/l^3+O(r^4)$.
%------------<fig>---------------------------
\begin{figure}[htbp]
\begin{center}
%\rotatebox{-90}{
\includegraphics[width=0.5\linewidth]{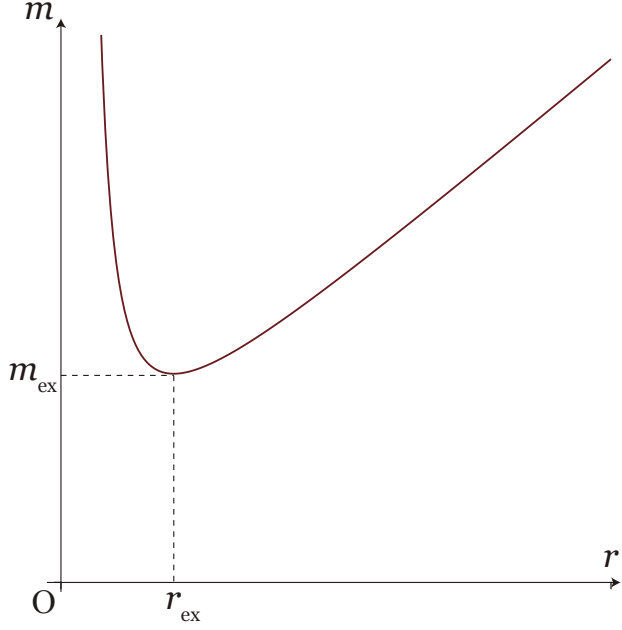}
%\subfigure[]{\includegraphics[width=0.7\linewidth]{Roberts-lambda1.eps}}
%\subfigure[]{\includegraphics[width=0.7\linewidth]{Roberts-lambda2.eps}}
%}
\caption{\label{M-rh-graph} The form of $m=m_{\rm h}(r)$ given by Eq.~(\ref{Bardeen-mh}).
}
\end{center}
\end{figure}
%--------------<fig>-----------------------

${\bar f}(r_{\rm h},m)(=f(r_{\rm h}))=0$ is solved for $m$ to give
\begin{align}
m=\frac{(r_{\rm h}^{2}+l^2)^{3/2}}{2r_{\rm h}^2}(=m_{\rm h}(r_{\rm h})), \label{Bardeen-mh}
\end{align}
which shows
\begin{align}
\frac{\D m_{\rm h}}{\D r_{\rm h}}=\frac{(r_{\rm h}^{2}-2l^2)\sqrt{r_{\rm h}^{2}+l^2}}{2r_{\rm h}^3}. \label{Bardeen-mh-d}
\end{align}
The form of the function $m_{\rm h}(r_{\rm h})$ is shown in Fig.~\ref{M-rh-graph} with $m_{\rm ex}=3\sqrt{3}l/4$ and $r_{\rm ex}=\sqrt{2}l$.
For $m>m_{\rm ex}$, the Bardeen spacetime represents a black hole with an outer horizon and an inner horizon located in the regions of $r>r_{\rm ex}$ and $r<r_{\rm ex}$, respectively.
For $m=m_{\rm ex}$, the spacetime becomes an extreme black hole with a degenerate horizon at $r=r_{\rm ex}$.
The Penrose diagrams of the Bardeen black hole are drawn in Fig.~\ref{PenroseSphericalNonsingularBH}.
For $0<m<m_{\rm ex}$ and $m<0$, the Bardeen spacetime represents a self-gravitating regular soliton without a horizon.

Equation~(\ref{matter-nonrot}) becomes
\begin{align}
\rho=-p_1=\frac{6ml^2}{(r^2+l^2)^{5/2}},\quad p_2=p_3=\frac{3ml^2(3r^2-2l^2)}{(r^2+l^2)^{7/2}},
\end{align} 
while Eq.~(\ref{matter-nonrotating}) gives
\begin{align}
&\rho+p_1=0,\quad \rho-p_1=\frac{12ml^2}{(r^2+l^2)^{5/2}},\\
&\rho+p_2=\rho+p_3=\frac{15ml^2r^2}{(r^2+l^2)^{7/2}},\\
&\rho-p_2=\rho-p_3=-\frac{3ml^2(r^2-4l^2)}{(r^2+l^2)^{7/2}},\\
&\rho+p_1+p_2+p_3=\frac{6ml^2(3r^2-2l^2)}{(r^2+l^2)^{7/2}},
\end{align} 
All the standard energy conditions are violated everywhere for $m<0$.
For $m>0$, while the WEC is satisfied everywhere, the DEC and SEC are satisfied only in the regions $0\le r\le 2l$ and $r\ge \sqrt{2/3}l$, respectively.
Since $m\ge m_{\rm ex}(>0)$ is required to be a black hole, the DEC is violated on the large event horizon satisfying $r_{\rm h}\ge 2l(> r_{\rm ex})$, while all the standard energy conditions are respected on the event horizon satisfying $r_{\rm ex}\le r_{\rm h}<2l$.

\subsubsection{Hayward spacetime}

The metric function $f(r)$ for the Hayward spacetime~\cite{Hayward2006} is 
\begin{align}
f(r)=1-\frac{2mr^2}{r^{3}+2ml^2}. \label{Hayward-BH-4-ap}
\end{align}
With $l\ne 0$ and $m\ge 0$, the spacetime is regular everywhere in the domain $r\in[0,\infty)$.
The metric function $f(r)$ behaves around $r=0$ as $f(r)\simeq 1-r^2/l^2+{O}(r^5)$ and therefore $r=0$ is regular but non-analytic.
With $l\ne 0$ and $m<0$, in contrast, there is a curvature singularity at $r=(-2ml^2)^{1/3}=:r_{\rm s}$, so that the domain of $r$ is $r\in(r_{\rm s},\infty)$ in this case.
%------------<fig>---------------------------
\begin{figure}[htbp]
\begin{center}
%\rotatebox{-90}{
\includegraphics[width=0.5\linewidth]{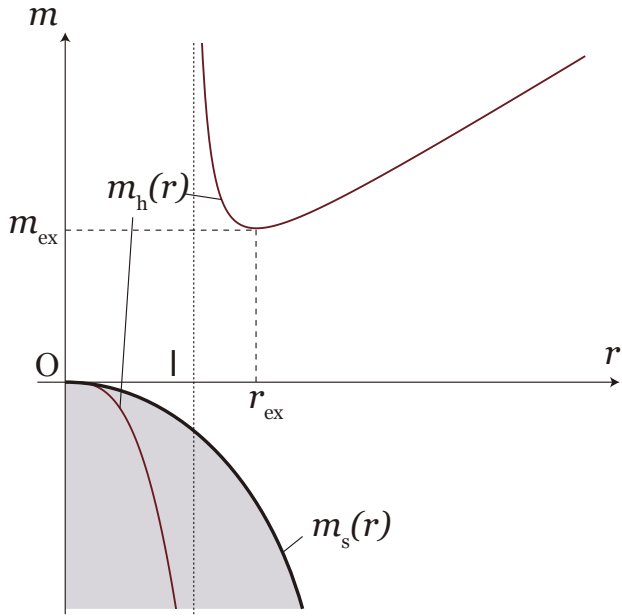}
%\subfigure[]{\includegraphics[width=0.7\linewidth]{Roberts-lambda1.eps}}
%\subfigure[]{\includegraphics[width=0.7\linewidth]{Roberts-lambda2.eps}}
%}
\caption{\label{M-rh-graph-H} The forms of $m=m_{\rm h}(r)$ (thin curves) and $m=m_{\rm s}(r)$ (a thick curve) given by Eq.~(\ref{Hayward-mh}) and~(\ref{Hayward-ms}), respectively.
}
\end{center}
\end{figure}
%--------------<fig>-----------------------

${\bar f}(r_{\rm h},m)(=f(r_{\rm h}))=0$ is solved for $m$ to give
\begin{align}
m=\frac{r_{\rm h}^3}{2(r_{\rm h}^2-l^2)}(=m_{\rm h}(r_{\rm h})), \label{Hayward-mh}
\end{align}
which shows
\begin{align}
\frac{\D m_{\rm h}}{\D r_{\rm h}}=\frac{r_{\rm h}^{2}(r_{\rm h}^{2}-3l^2)}{2(r_{\rm h}^2-l^2)^2}. \label{Hayward-mh-d}
\end{align}
The relation between the mass parameter $m$ and the singularity radius $r_{\rm s}$ for $m<0$ is given by 
\begin{align}
m=-\frac{r_{\rm s}^3}{2l^2}=:m_{\rm s}(r_{\rm s}). \label{Hayward-ms}
\end{align}
By the following expression
\begin{align}
m_{\rm s}(r)-m_{\rm h}(r)=\frac{r^5}{2l^2(l^2-r^2)},
\end{align}
$m_{\rm s}(r)>(<)m_{\rm h}(r)$ is satisfied in the domain $r<(>)l$.
The forms of the functions $m_{\rm h}(r)$ and $m_{\rm s}(r)$ are shown in Fig.~\ref{M-rh-graph-H} with $m_{\rm ex}=3\sqrt{3}l/4$ and $r_{\rm ex}=\sqrt{3}l$.
The spacetimes represented in the regions $m>m_{\rm s}(r)$ (unshaded) and $m<m_{\rm s}(r)$ (shaded) are distinct.
We are interested in the region of $m>m_{\rm s}(r)$ representing an asymptotically flat spacetime.

For $m>m_{\rm ex}$, the Hayward spacetime represents a black hole with an outer horizon and an inner horizon located in the regions of $r>r_{\rm ex}$ and $r<r_{\rm ex}$, respectively.
For $m=m_{\rm ex}$, the spacetime becomes an extreme black hole with a degenerate horizon at $r=r_{\rm ex}$.
The Penrose diagrams of the Hayward black hole are the same as those in Fig.~\ref{PenroseSphericalNonsingularBH}.
For $0<m<m_{\rm ex}$, the spacetime represents a self-gravitating regular soliton without a horizon.
For $m<0$, the spacetime represents a naked singularity located at a finite radius $r=r_{\rm s}:=(-2ml^2)^{1/3}(>0)$.

Equation~(\ref{matter-nonrot}) becomes
\begin{align}
\rho=-p_1=\frac{12m^2l^2}{(r^3+2ml^2)^{2}},\quad p_2=p_3=\frac{24m^2l^2(r^3-ml^2)}{(r^3+2ml^2)^{3}},
\end{align} 
while Eq.~(\ref{matter-nonrotating}) gives
\begin{align}
&\rho+p_1=0,\quad \rho-p_1=\frac{24m^2l^2}{(r^3+2ml^2)^{2}},\\
&\rho+p_2=\rho+p_3=\frac{36m^2l^2r^3}{(r^3+2ml^2)^{3}},\\
&\rho-p_2=\rho-p_3=-\frac{12m^2l^2(r^3-4ml^2)}{(r^3+2ml^2)^{3}},\label{Hayward-EC-1}\\
&\rho+p_1+2p_2=\frac{48m^2l^2(r^3-ml^2)}{(r^3+2ml^2)^{3}}.\label{Hayward-EC-2}
\end{align} 
Hence, the WEC (and the NEC as well) is respected everywhere for any $m$.
For $m>0$, the DEC and SEC are respected only in the regions of $0\le r\le (4ml^2)^{1/3}$ and $r\ge (ml^2)^{1/3}$, respectively.
For $m<0$, the DEC is violated and the SEC is respected in the whole domain $r\in(r_{\rm s},\infty)$.

Substituting $m=m_{\rm h}(r_{\rm h})$ given by Eq.~(\ref{Hayward-mh}) into Eqs.~(\ref{Hayward-EC-1}) and (\ref{Hayward-EC-2}), we obtain
\begin{align}
&(\rho-p_2)|_{r=r_{\rm h}}=(\rho-p_3)|_{r=r_{\rm h}}=\frac{3l^2(3l^2-r_{\rm h}^2)}{r_{\rm h}^6},\\
&(\rho+p_1+2p_2)|_{r=r_{\rm h}}=\frac{6l^2(2r_{\rm h}^2-3l^2)}{r_{\rm h}^6}.
\end{align} 
Accordingly, the DEC is violated on the non-degenerate event horizon since $r_{\rm h}>r_{\rm ex}(=\sqrt{3}l)$ is satisfied. On the degenerate event horizon ($r_{\rm h}=r_{\rm ex}$), all the standard energy conditions are respected.

\subsubsection{Dymnikova spacetime}

Dymnikova presented the following metric function $f(r)$ as a solution with a nonlinear electromagnetic field in general relativity~\cite{Dymnikova:2004zc};
\begin{align}
f(r)=1-\frac{4m}{\pi r}\biggl\{\arctan\biggl(\frac{r}{l}\biggl)-\frac{lr}{r^2+l^2}\biggl\}, \label{nonsingular-D}
\end{align}
where we have reparametrized the solution.
The metric is invariant for $r\to -r$ and the spacetime is analytic everywhere in the domain $r\in[0,\infty)$.
The metric function $f(r)$ behaves around $r=0$ as $f(r)\simeq 1-8mr^2/(3\pi l^3)+{O}(r^4)$.

${\bar f}(r_{\rm h},m)(=f(r_{\rm h}))=0$ is solved for $m$ to give
\begin{align}
m=\frac{\pi r_{\rm h}}{4}\biggl\{\arctan\biggl(\frac{r_{\rm h}}{l}\biggl)-\frac{lr_{\rm h}}{r_{\rm h}^2+l^2}\biggl\}^{-1}(=m_{\rm h}(r_{\rm h})), \label{D-mh}
\end{align}
which shows
\begin{align}
\frac{\D m_{\rm h}}{\D r_{\rm h}}=\frac{\pi \{(r_{\rm h}^2+l^2)^2\arctan(r_{\rm h}/l)-lr_{\rm h}(3r_{\rm h}^2+l^2)\}}{4\{(r_{\rm h}^2+l^2)\arctan(r_{\rm h}/l)-lr_{\rm h}\}^2}. \label{D-mh-d}
\end{align}
The function $m_{\rm h}(r_{\rm h})$ has a single positive local minimum at $r_{\rm h}=r_{\rm ex}(\simeq 1.83l)$ and it obeys $m_{\rm h}(r_{\rm h})\to \infty$ as $r_{\rm h}\to 0$ and $r_{\rm h}\to \infty$.
Positivity of $m_{\rm h}(r_{\rm ex})$ is shown as
\begin{align}
m_{\rm h}(r_{\rm ex})=\frac{\pi(r_{\rm ex}^2+l^2)^2}{8lr_{\rm ex}^2}=:m_{\rm ex}(\simeq 2.21l).
\end{align}
Thus, the form of the function $m_{\rm h}(r_{\rm h})$ is the same as Fig.~\ref{M-rh-graph} but the values of $m_{\rm ex}$ and $r_{\rm ex}$ are different.
For $m>m_{\rm ex}$, Dymnikova spacetime represents a black hole with an outer horizon and an inner horizon located in the regions of $r>r_{\rm ex}$ and $r<r_{\rm ex}$, respectively.
For $m=m_{\rm ex}$, it becomes an extreme black hole with one degenerate horizon at $r=r_{\rm ex}$.
The Penrose diagrams of the Dymnikova black hole are the same as those in Fig.~\ref{PenroseSphericalNonsingularBH}.
For $0<m<m_{\rm ex}$ and $m<0$, the spacetime represents a self-gravitating regular soliton without a horizon.

Equation~(\ref{matter-nonrot}) becomes
\begin{align}
\rho=-p_1=\frac{8ml}{\pi(r^2+l^2)^2},\quad p_2=p_2=\frac{8ml(r^2-l^2)}{\pi(r^2+l^2)^3}.
\end{align} 
while Eq.~(\ref{matter-nonrotating}) shows
\begin{align}
&\rho+p_1=0,\quad \rho-p_1=\frac{16ml}{\pi(r^2+l^2)^2},\\
&\rho+p_2=\rho+p_3=\frac{16mlr^2}{\pi(r^2+l^2)^3},\\
&\rho-p_2=\rho-p_3=\frac{16ml^3}{\pi(r^2+l^2)^3},\\
&\rho+p_1+p_2+p_2=\frac{16ml(r^2-l^2)}{\pi(r^2+l^2)^3}.
\end{align} 
All the standard energy conditions are violated everywhere for $m<0$.
For $m>0$, the DEC (and hence the WEC and NEC as well) is satisfied everywhere, while the SEC are satisfied only in the region of $r\ge l$.
On the event horizon of the Dymnikova black hole, all the standard energy conditions are respected since $m\ge m_{\rm ex}(>0)$ and $r_{\rm h}\ge r_{\rm ex}(\simeq 1.83l)$ are satisfied.

\subsubsection{Fan-Wang spacetime}

Fan and Wang derived the following metric function $f(r)$ as a solution with a nonlinear electromagnetic field in general relativity~\cite{Fan:2016hvf}:
\begin{align}
f(r)=1-\frac{2mr^2}{(r+l)^3}.\label{new-nonsingular-ap}
\end{align}
In fact, their metric function $f(r)$ is more general and contains a term $-2{\bar M}/r$, where ${\bar M}$ is an integration constant.
However, since a regular center is achieved only for ${\bar M}=0$, we refer to the spacetime with the metric function (\ref{new-nonsingular-ap}) as the ``Fan-Wang spacetime'' in the present paper.
With $l\ne 0$, the Fan-Wang spacetime is regular everywhere in the domain $r\in[0,\infty)$.
The metric function $f(r)$ behaves around $r=0$ as $f(r)\simeq 1-2mr^2/l^3+{O}(r^3)$ and therefore $r=0$ is regular but non-analytic.

For the Fan-Wang spacetime, ${\bar f}(r_{\rm h},m)(=f(r_{\rm h}))=0$ is solved for $m$ to give
\begin{align}
m=\frac{(r_{\rm h}+l)^3}{2r_{\rm h}^2}(=m_{\rm h}(r_{\rm h})), \label{new-mh}
\end{align}
which shows
\begin{align}
\frac{\D m_{\rm h}}{\D r_{\rm h}}=\frac{(r_{\rm h}-2l)(r_{\rm h}+l)^2}{2r_{\rm h}^3}.\label{new-mh-d}
\end{align}
The form of the function $m_{\rm h}(r)$ is the same as Fig.~\ref{M-rh-graph-H} with $m_{\rm ex}=27l/8$ and $r_{\rm ex}=2l$.
For $m>m_{\rm ex}$, the Fan-Wang spacetime represents a black hole with an outer horizon and an inner horizon located in the regions of $r>r_{\rm ex}$ and $r<r_{\rm ex}$, respectively.
For $m=m_{\rm ex}$, the spacetime becomes an extreme black hole with a degenerate horizon at $r=r_{\rm ex}$.
The Penrose diagrams of the Fan-Wang black hole are the same as those in Fig.~\ref{PenroseSphericalNonsingularBH}.
For $0<m<m_{\rm ex}$ and $m<0$, the Fan-Wang spacetime represents a self-gravitating regular soliton without a horizon.

Equation~(\ref{matter-nonrot}) becomes
\begin{align}
\rho=-p_1=\frac{6ml}{(r+l)^4},\quad p_2=p_3=\frac{6ml(r-l)}{(r+l)^5}
\end{align} 
and Eq.~(\ref{matter-nonrotating}) gives
\begin{align}
&\rho+p_1=0,\quad \rho-p_1=\frac{12ml}{(r+l)^4},\\
&\rho+p_2=\rho+p_3=\frac{12mlr}{(r+l)^5},\\
&\rho-p_2=\rho-p_3=\frac{12ml^2}{(r+l)^5},\\
&\rho+p_1+2p_2=\frac{12ml(r-l)}{(r+l)^5}.
\end{align} 
All the standard energy conditions are violated everywhere for $m<0$.
For $m>0$, the DEC is respected everywhere (and hence the WEC and NEC as well), while the SEC are respected only in the region of $r\ge l$.
On the event horizon of the Fan-Wang black hole, all the standard energy conditions are respected since $m\ge m_{\rm ex}(>0)$ and $r_{\rm h}\ge r_{\rm ex}(=2l)$ are satisfied.
%-------------- TABLE ---------------%
\begin{table*}[htb]
\begin{center}
\caption{\label{table:BH} The domains of $r$ where the standard energy conditions are respected in the Bardeen, Hayward, Dymnikova, and Fan-Wang spacetimes with $l> 0$.}
\scalebox{0.8}{
\begin{tabular}{|c|c|c|c|c||c|c|}
\hline
& NEC & WEC & DEC & SEC & $r=0$ & Black Hole \\ \hline\hline
Bardeen ($m>0$) & everywhere & everywhere & $0\le r\le 2l$ & $r\ge \sqrt{2/3}l$ & Analytic & $m\ge 3\sqrt{3}l/4$ \\ \hline
Bardeen ($m<0$) & $\emptyset$ & $\emptyset$ & $\emptyset$ & $\emptyset$ & Analytic & n/a \\ \hline
Hayward ($m>0$) & everywhere & everywhere & $0\le r\le (4ml^2)^{1/3}$ & $r\ge (ml^2)^{1/3}$ & Regular & $m\ge 3\sqrt{3}l/4$ \\ \hline
Hayward ($m<0$) & everywhere & everywhere & $\emptyset$ & everywhere & n/a & n/a \\ \hline
Dymnikova ($m>0$) & everywhere & everywhere & everywhere & $r\ge l$ & Analytic & $m\gtrsim 2.21l$ \\ \hline
Dymnikova ($m<0$) & $\emptyset$ & $\emptyset$ & $\emptyset$ & $\emptyset$ & Analytic & n/a \\ \hline
Fan-Wang ($m>0$) & everywhere & everywhere & everywhere & $r\ge l$ & Regular & $m\ge 27l/8$ \\ \hline
Fan-Wang ($m<0$) & $\emptyset$ & $\emptyset$ & $\emptyset$ & $\emptyset$ & Regular & n/a \\ \hline
\end{tabular} 
}
\end{center}
\end{table*}
%------------------------------------%
%-------------- TABLE ---------------%
\begin{table*}[htb]
\begin{center}
\caption{\label{table:summary} Results of the geometric criteria C1--C5 in Sec.~\ref{sec:intro} for spherically symmetric non-singular black holes.}
\scalebox{1.0}{
\begin{tabular}{|c|c|c|c|c|c|}
\hline
 & C1 & C2 & C3 & C4 & C5 \\ \hline\hline
Bardeen & \checkmark  & \checkmark & $\times$ & $\times$ & $\times$ \\ \hline
Hayward & \checkmark  & \checkmark & $\times$ & $\times$ & \checkmark \\ \hline
Dymnikova & \checkmark  & \checkmark & \checkmark & \checkmark & $\times$ \\ \hline
Fan-Wang & \checkmark  & \checkmark & \checkmark & \checkmark & $\times$ \\ \hline
\end{tabular} 
}
\end{center}
\end{table*}
%------------------------------------%

Our results in the spherically symmetric case are summarized in Tables~\ref{table:BH} and \ref{table:summary}.
The SEC is violated around the regular center $r=0$ in all the non-singular black-hole spacetimes, which means that the gravitational force becomes repulsive there.
Although the WEC is respected everywhere, the Bardeen and Hayward black-hole spacetimes violate the DEC in the asymptotically flat region, so that they do not satisfy the criterion C5 in Sec.~\ref{sec:intro}.
In contrast, the Dymnikova and Fan-Wang black-hole spacetimes respect the DEC everywhere.
Figure~\ref{PenroseSphericalNonsingularBH} shows the Penrose diagrams of all these four black holes.

\subsection{Rotating counterparts}

Now let us study the rotating counterparts of the Bardeen, Hayward, Dymnikova, and Fan-Wang spacetimes described by the metric (\ref{BL}) with the metric functions (\ref{m-B})--(\ref{m-FW}), respectively. 
By Proposition~\ref{Prop:conical} with Eqs.~(\ref{m-B-r=0})--(\ref{m-FW-r=0}), these spacetimes are analytic at $r=0$ with $\theta\ne \pi/2$ and can be extended beyond $r=0$ into the region with negative $r$.
Then, properties of the rotating Bardeen and Dymnikova spacetimes in the regions with $r>0$ and $r<0$ are the same because the G\"urses-G\"ursey metric (\ref{BL}) is invariant for $r\to -r$ and $a\to -a$ if $M(r)$ is an odd function.

\subsubsection{Singularities, stable causality, and energy conditions}

%\noindent
%{\bf Scalar polynomial curvature singularities}

By Proposition~\ref{Prop:conical} with Eqs.~(\ref{m-B-r=0})--(\ref{m-FW-r=0}), $(r,\theta)=(0,\pi/2)$ is a ring-like conical singularity and not a scalar polynomial curvature singularity.
However, the rotating Hayward and Fan-Wang spacetimes with $m\ne 0$ and $l>0$ contain a scalar polynomial curvature singularity in the region with negative $r$.
The Ricci scalar $R$ of the rotating Fan-Wang spacetime is given by 
\begin{align}
R=\frac{24ml^2r^2}{(r+l)^5(r^2+a^2\cos^2\theta)},
\end{align}
which blows up at $r= -l(<0)$.
The Ricci scalar of the rotating Hayward spacetime is 
\begin{align}
R=-\frac{24m^2l^2r^2(r^3-4ml^2)}{(r^3+2ml^2)^3(r^2+a^2\cos^2\theta)},
\end{align}
which blows up at $r= -\epsilon(2|m|l^2)^{1/3}$, where $\epsilon=\pm 1$ is the sign of $m$.

In contrast, the rotating Bardeen and Dymnikova spacetimes with $m\ne 0$ and $l> 0$ are regular everywhere in the domain $r\in(-\infty,\infty)$ other than $(r,\theta)=(0,\pi/2)$.
The metric functions $\Delta(r)$ with the mass functions (\ref{m-B}) and (\ref{m-D}) are given by 
\begin{align}
\mbox{Bardeen}:~&\Delta(r)=r^2+a^2-\frac{2mr^4}{(r^{2}+l^2)^{3/2}},\label{Delta-B}\\
\mbox{Dymnikova}:~&\Delta(r)=r^2+a^2-\frac{4mr}{\pi}\biggl\{\arctan\biggl(\frac{r}{l}\biggl)-\frac{lr}{r^2+l^2}\biggl\}\label{Delta-D}
\end{align}
and hence the metric is regular everywhere except at $(r,\theta)=(0,\pi/2)$.

%\noindent
%{\bf Stable causality}

In the Kerr spacetime described by the metric (\ref{BL}) with $M(r)=m(>0)$, there is a closed timelike curve with its tangent vector $u^\mu(\partial/\partial x^\mu)=\partial/\partial \phi$ in the negative region of $r$ near $(r,\theta)=(0,\pi/2)$, where $g_{\phi\phi}$ becomes negative.
In contrast, the rotating Bardeen, Hayward, Dymnikova, and Fan-Wang spacetimes with $m>0$ are stably causal and therefore do not contain closed causal curves.
Equations~(\ref{m-D})--(\ref{m-FW}) give
\begin{align}
\mbox{Bardeen}:~&rM(r)=\frac{mr^4}{(r^{2}+l^2)^{3/2}},\label{m-B-r}\\
\mbox{Hayward}:~&rM(r)=\frac{mr^4}{r^{3}+2ml^2}, \label{m-Hayward-r}\\
\mbox{Dymnikova}:~&rM(r)=\frac{2mr}{\pi}\biggl\{\arctan\biggl(\frac{r}{l}\biggl)-\frac{lr}{r^2+l^2}\biggl\}, \label{m-D-r}\\
\mbox{Fan-Wang}:~&rM(r)=\frac{mr^4}{(r+l)^3}.\label{m-FW-r}
\end{align}
The domains of $r$ are $r\in(-\infty,\infty)$ in the rotating Bardeen and Dymnikova spacetimes and $r\in(r_{\rm s},\infty)$ in the rotating Hayward and Fan-Wang spacetimes, where $r=r_{\rm s}(<0)$ is the singularity radius.
Since $rM(r)\ge 0$ holds in these domains for $m>0$, the spacetimes are stably causal by Proposition~\ref{Prop:Causal}.

%\noindent
%{\bf Energy conditions}

Let us check the standard energy conditions in the asymptotically flat region $r\to \infty$.
Since the numerators of $\rho$ in Eq.~(\ref{matter-rot}) and of $\rho-p_1$ and $\rho-p_2(=\rho-p_3)$ in Eq.~(\ref{matter-rotating}) do not contain $a$, their leading orders near $r\to \infty$ are the same as those in the spherically symmetric case given in Eq.~(\ref{EC-infty-expand}).
On the other hand, under the asymptotic behavior $M(r)\simeq M_0+M_1/r^\beta$ with $\beta>0$ and $M_0M_1\ne 0$ near $r\to \infty$, which includes the cases of Eqs.~(\ref{m-B-r=infty})--(\ref{m-FW-r=infty}), Eq.~(\ref{matter-rotating}) gives
\begin{align}
\begin{aligned}
&\rho+p_2(=\rho+p_3)\simeq -\frac{\beta(\beta+3)M_1}{r^{\beta+3}},\\
&\rho+p_1+p_2+p_3\simeq -\frac{2\beta(\beta+1)M_1}{r^{\beta+3}},
\end{aligned}
\end{align}
which do not contain $a$.
Accordingly, the energy conditions near $r\to \infty$ in the rotating case are understood in Table~\ref{table:BH} in the spherically symmetric case.
Thus, all the standard energy conditions are satisfied near $r\to \infty$ only in the rotating Dymnikova and Fan-Wang spacetimes with $m>0$.

In the spacetime (\ref{BL}) with the metric functions (\ref{m-B})--(\ref{m-FW}), the equation $\Delta(r_{\rm h})=0$ to determine the location of a Killing horizon $r=r_{\rm h}$ is solved to define a function $m=m_{\rm h}(r_{\rm h})$.
Substituting $m=m_{\rm h}(r_{\rm h})$ into Eqs.~(\ref{matter-rot}) and (\ref{matter-rotating}) and taking the large-horizon limit $r_{\rm h}\to \infty$, we obtain the leading terms of the quantities relevant to the energy conditions as shown in Table~\ref{table:EC-horizon-infty}.
Hence, all the standard energy conditions are satisfied on the event horizon of the large rotating Dymnikova and Fan-Wang black holes.
In contrast, the DEC is violated on the event horizon of the large rotating Bardeen and Hayward black holes and therefore they do not satisfy the criterion C4 in Sec.~\ref{sec:intro}.
%-------------- TABLE ---------------%
\begin{table*}[htb]
\begin{center}
\caption{\label{table:EC-horizon-infty} The leading terms in the limit $r_{\rm h}\to \infty$ of the quantities in Eqs.~(\ref{matter-rot}) and (\ref{matter-rotating}) evaluated on a Killing horizon $r=r_{\rm h}$ in the rotating non-singular black holes with $l> 0$.}
\scalebox{0.95}{
\begin{tabular}{|c|c|c|c|c|c|c|}
\hline
& $\rho$ & $\rho+p_1$ & $\rho-p_1$ & $\rho+p_2$ & $\rho-p_2$ & $\rho+p_1+2p_2$ \\ \hline\hline
Rotating Bardeen & $3l^2/r_{\rm h}^4$ & $0$ & $6l^2/r_{\rm h}^4$ & $15l^2/(2r_{\rm h}^4)$ & $-3l^2/(2r_{\rm h}^4)$ & $9l^2/r_{\rm h}^4$ \\ \hline
Rotating Hayward & $3l^2/r_{\rm h}^4$ & $0$ & $6l^2/r_{\rm h}^4$ & $9l^2/r_{\rm h}^4$ & $-3l^2/r_{\rm h}^4$ & $12l^2/r_{\rm h}^4$ \\ \hline
Rotating Dymnikova & $4l/(\pi r_{\rm h}^3)$ & $0$ & $8l/(\pi r_{\rm h}^3)$ & $8l/(\pi r_{\rm h}^3)$ & $8l^3/(\pi r_{\rm h}^5)$ & $8l/(\pi r_{\rm h}^3)$ \\ \hline
Rotating Fan-Wang & $3l/r_{\rm h}^3$ & $0$ & $6l/r_{\rm h}^3$ & $6l/r_{\rm h}^3$ & $6l^2/r_{\rm h}^4$ & $6l/r_{\rm h}^3$ \\ \hline
\end{tabular} 
}
\end{center}
\end{table*}
%------------------------------------%

%-------------- TABLE ---------------%
\begin{table*}[htb]
\begin{center}
\caption{\label{table:summary-rotating} Results of geometric criteria C1--C5 in Sec.~\ref{sec:intro} for rotating non-singular black holes with $l> 0$. For the rotating Bardeen and Dymnikova black holes, the criterion W-C1 is satisfied if $(r,\theta)=(0,\pi/2)$ is not a p.p. curvature singularity.}
\scalebox{1.0}{
\begin{tabular}{|c|c|c|c|c|c|c|}
\hline
 & C1 & W-C1 & C2 & C3 & C4 & C5 \\ \hline\hline
Rotating Bardeen & $\times$ & See the caption & \checkmark & $\times$ & $\times$ & $\times$ \\ \hline
Rotating Hayward & $\times$ & $\times$ & \checkmark & $\times$ & $\times$ & \checkmark \\ \hline
Rotating Dymnikova & $\times$ & See the caption & \checkmark & \checkmark & \checkmark & $\times$ \\ \hline
Rotating Fan-Wang & $\times$ & $\times$ & \checkmark & \checkmark & \checkmark & $\times$ \\ \hline
\end{tabular} 
}
\end{center}
\end{table*}
%------------------------------------%

\subsubsection{More on the rotating Dymnikova black hole}

Our results in the rotating case are summarized in Table~\ref{table:summary-rotating}, which indicates that the rotating Dymnikova black hole is the most preferable among the four. 
Now let us study more on the rotating Dymnikova spacetime, described by the metric (\ref{BL}) with the mass function (\ref{m-D}).
Since the mass function $M(r)$ is odd, properties of the spacetime in the regions with $r>0$ and $r<0$ are the same and therefore $r\to-\infty$ is another asymptotically flat region.
Hereafter we consider the domain $r\in [0,\infty)$.

In order to clarify the global structure of the spacetime (\ref{BL}), the following lemma is useful, which is a straightforward generalization of Lemma~\ref{lm:horizon-nonrotating}.
%----------------------- lemma ------------------------------%
\begin{lm}
\label{lm:horizon-rotating}
Identify the metric function $\Delta(r)$ in the spacetime (\ref{BL}) as a function ${\bar \Delta}(r,m)$ of $r$ and $m$ such that ${\bar \Delta}(r,m)\equiv \Delta(r)$.
Define a function $m=m_{\rm h}(r_{\rm h})$ as a solution of ${\bar \Delta}(r_{\rm h},m)=0$, where $r=r_{\rm h}(>0)$ is the radius of a Killing horizon, and suppose that $\partial{\bar \Delta}/\partial m|_{r=r_{\rm h}}$ is non-zero and finite.
Then, in the case of $\partial{\bar \Delta}/\partial m|_{r=r_{\rm h}}<(>)0$, a Killing horizon is outer if $\D m_{\rm h}/\D r_{\rm h}>(<)0$, inner if $\D m_{\rm h}/\D r_{\rm h}<(>)0$, and degenerate if $\D m_{\rm h}/\D r_{\rm h}=0$.
\end{lm}
%----------------------- lemma ------------------------------%

\noindent
For the rotating Dymnikova spacetime, $\partial{\bar \Delta}/\partial m<0$ holds in the domain $r\in(0,\infty)$ because $\partial{\bar \Delta}/\partial m=r^2\partial{\bar f}/\partial m$ and $\partial{\bar f}/\partial m<0$ are satisfied.

\begin{figure}[htbp]
\begin{center}
%\rotatebox{-90}{
\includegraphics[width=0.5\linewidth]{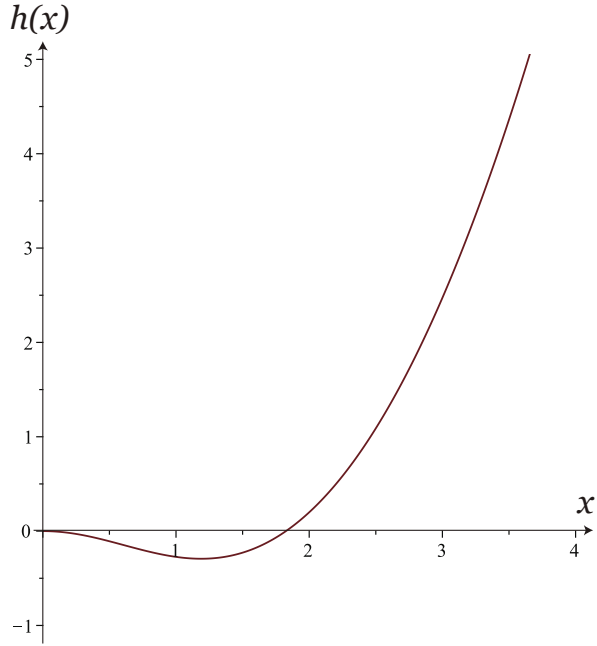}
%\subfigure[]{\includegraphics[width=0.7\linewidth]{Roberts-lambda1.eps}}
%\subfigure[]{\includegraphics[width=0.7\linewidth]{Roberts-lambda2.eps}}
%}
\caption{\label{hx-graph} The form of the function $h(x)$ defined by Eq.~(\ref{D-m'}).
}
\end{center}
\end{figure}

Now we show that there is a positive lower bound of $m$ for a black-hole configuration of the rotating Dymnikova spacetime with $l>0$ and the rotating Dymnikova black hole admits two Killing horizons at most in the region with $r> 0$.
By Eq.~(\ref{Delta-D}), ${\bar \Delta}(r,m)=0$ is solved for $m$ to give
\begin{align}
m=\frac{\pi(r_{\rm h}^2+a^2)}{4r_{\rm h}}\biggl\{\arctan\biggl(\frac{r_{\rm h}}{l}\biggl)-\frac{lr_{\rm h}}{r_{\rm h}^2+l^2}\biggl\}^{-1}(=m_{\rm h}(r_{\rm h})),\label{M-rh-D-rot}
\end{align}
which shows
\begin{align}
m_{\rm h}'(r_{\rm h})=&\frac{\pi w(r_{\rm h})}{4r_{\rm h}^2(r_{\rm h}^2+l^2)^2}\biggl\{\arctan\biggl(\frac{r_{\rm h}}{l}\biggl)-\frac{lr_{\rm h}}{r_{\rm h}^2+l^2}\biggl\}^{-2},\label{mh'-D}
\end{align}
where $w(r_{\rm h})$ is defined by 
\begin{align}
w(r_{\rm h}):=(r_{\rm h}^2+l^2)^2(r_{\rm h}^2-a^2)\arctan\biggl(\frac{r_{\rm h}}{l}\biggl) -lr_{\rm h}\left\{3r_{\rm h}^4+(l^2+a^2)r_{\rm h}^2-a^2l^2\right\}.\label{w}
\end{align}
Equation~(\ref{M-rh-D-rot}) shows that $m(r_{\rm h})\to \infty$ holds as $r_{\rm h}\to 0$ and $r_{\rm h}\to \infty$.
On the other hand, $w(r_{\rm h})=0$, which is equivalent to $m_{\rm h}'(r_{\rm h})=0$, gives
\begin{align}
\frac{a^2}{l^2}=\frac{x^2\{(1+x^2)^2\arctan x-x(1+3x^2)\}}{(1+x^2)^2\arctan x-x(1-x^2)}=:h(x),\label{D-m'}
\end{align}
where $x:=r_{\rm h}/l$.
As plotted in Fig.~\ref{hx-graph}, the function $h(x)$ is increasing if it is positive and therefore $m_{\rm h}'(r_{\rm h})=0$ admits a single real solution at $r_{\rm h}=r_{\rm ex}(\gtrsim 1.83l)$ in the domain of $r_{\rm h}\ge 0$.
Thus, $m_{\rm h}(r_{\rm h})$ has a single local minimum at $r_{\rm h}=r_{\rm ex}$ and $m_{\rm h}(r_{\rm h})$ is decreasing (increasing) in the region of $r_{\rm h}<(>)r_{\rm ex}$.
The minimum value of $m_{\rm h}(r_{\rm h})$ is given by 
\begin{align}
m_{\rm h}(r_{\rm ex})=\frac{\pi(r_{\rm ex}^2-a^2)(r_{\rm ex}^2+l^2)^2}{8lr_{\rm ex}^4}=:m_{\rm ex}. \label{mex}
\end{align}
Equation~(\ref{w}) gives $w(|a|)=-4l|a|^5(<0)$, which shows $m_{\rm h}'(|a|)<0$ by Eq.~(\ref{mh'-D}) and hence $r_{\rm ex}>|a|$ is satisfied.
Therefore, the constant $m_{\rm ex}$ defined by Eq.~(\ref{mex}) is positive, so that the shape of the function $m_{\rm h}(r_{\rm h})$ is the same as Fig.~\ref{M-rh-graph}.
%------------<fig>---------------------------
%--------------<fig>-----------------------

\begin{figure}[htbp]
\begin{center}
%\rotatebox{-90}{
\includegraphics[width=1.0\linewidth]{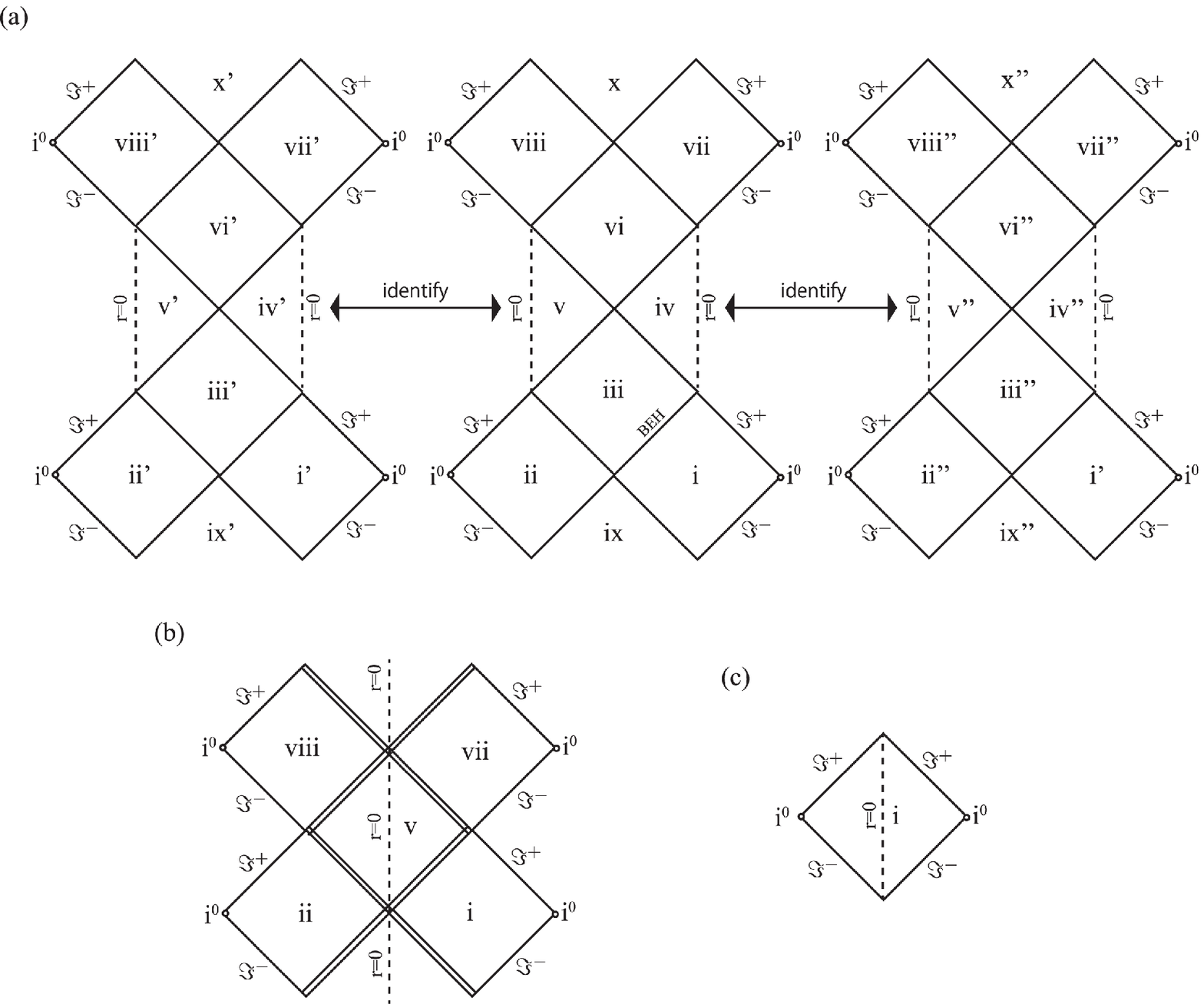}
%\subfigure[]{\includegraphics[width=0.7\linewidth]{Roberts-lambda1.eps}}
%\subfigure[]{\includegraphics[width=0.7\linewidth]{Roberts-lambda2.eps}}
%}
\caption{\label{PenroseRotatingDymnikova} Penrose diagrams of the rotating Dymnikova black hole with (a) non-degenerate horizons ($m>m_{\rm ex}$), (b) degenerate horizons ($m=m_{\rm ex}$), and (c) no horizon ($m<m_{\rm ex}$).
A conical singularity at $(r,\theta)=(0,\pi/2)$ is not a scalar polynomial curvature singularity.
}
\end{center}
\end{figure}

As a result, in the domain $r\in[0,\infty)$ of the rotating Dymnikova spacetime with $m> m_{\rm ex}(>0)$, there are a black-hole event horizon at $r=r_2(>r_{\rm ex})$ and an inner horizon at $r=r_1$ satisfying $0<r_1<r_{\rm ex}$. 
Since the spacetime structures in the regions of $r>0$ and $r<0$ are the same, there are four Killing horizons in total in the whole domain $r\in(-\infty,\infty)$.
For $m= m_{\rm ex}$, the spacetime represents an extreme black hole with two degenerate Killing horizons at $r=r_{\rm ex}$ and $r=-r_{\rm ex}$.
For $0<m< m_{\rm ex}$ and $m<0$, the spacetime does not admit a Killing horizon and represents a rotating self-gravitating soliton with a conical singularity at $(r,\theta)=(0,\pi/2)$.
The Penrose diagrams of the rotating Dymnikova spacetime are drawn in Fig.~\ref{PenroseRotatingDymnikova}.
In all the cases, the spacetime has a wormhole structure, of which throat is located at $r=0$.
%------------<fig>---------------------------
%--------------<fig>-----------------------

Unfortunately, by Proposition~\ref{Prop:no-go-rot} and Eq.~(\ref{m-D-r=0}), all the standard energy conditions are violated around $r=0$ with $\theta\ne \pi/2$ in the rotating Dymnikova black-hole spacetime.
In contrast, on the equatorial plane $\theta=\pi/2$, Eqs.~(\ref{matter-rot}) and (\ref{matter-rotating}) are identical to Eqs.~(\ref{matter-nonrot}) and (\ref{matter-nonrotating}) in the spherically symmetric case and therefore only the SEC is violated in the region with $-l<r<l$ as shown by Table~\ref{table:BH}.
Nevertheless, we show that the DEC is respected outside the event horizons in the rotating Dymnikova black-hole spacetime realized for $m\ge m_{\rm ex}$.
More precisely, we show that the DEC is respected in the domain $r\in [r_0,\infty)$, where $r_0:=\sqrt{|al|}$ satisfies $r_0<r_{\rm ex}$.

Since the spacetime is invariant for $(r,a)\to (-r,-a)$, we study the domain $r\in [0,\infty)$.
Equations (\ref{matter-rot}) and (\ref{matter-rotating}) with the mass function (\ref{m-D}) give
\begin{align}
&\rho=\frac{8mlr^4}{\pi(r^2+l^2)^2\Sigma^2},\quad \rho+p_1=0,\qquad \rho-p_1=\frac{16mlr^4}{\pi(r^2+l^2)^2\Sigma^2},\\
&\rho+p_2=\rho+p_3=\frac{16mlr^2(r^2+al\cos\theta)(r^2-al\cos\theta)}{\pi(r^2+l^2)^3\Sigma^2},\\
&\rho-p_2=\rho-p_3=\frac{16ml^3r^2}{\pi(r^2+l^2)^3\Sigma},\\
&\rho+p_1+p_2+p_3=\frac{16mlr^2(r^4-l^2r^2-2l^2a^2\cos^2\theta)}{\pi(r^2+l^2)^3\Sigma^2}.\label{SEC-r-D}
\end{align}
By Eqs.~(\ref{NEC})--(\ref{DEC}), the DEC is respected with $m\ge 0$ and $l> 0$ in the region $r^2\ge r_{\rm DEC}^2(\theta)$, where
\begin{align}
r_{\rm DEC}(\theta):=\sqrt{l|a\cos\theta|}.
\end{align}
We define the upper bound of $r_{\rm DEC}(\theta)$ as $r_0:=\sqrt{|al|}(\ge r_{\rm DEC}(\theta))$.
Since Eq.~(\ref{w}) gives
\begin{align}
w(r_0)=&|a|l^3(l+|a|)^2\biggl\{\biggl(1-\frac{|a|}{l}\biggl)\arctan\biggl(\sqrt{\frac{|a|}{l}}\biggl)-\sqrt{\frac{|a|}{l}}\biggl\}<0,
\end{align}
we obtain $m_{\rm h}'(r_0)<0$ by Eq.~(\ref{mh'-D}), which implies $r_{\rm DEC}(\theta)\le r_0<r_{\rm ex}$.
Since the radius of the event horizon $r=r_2$ satisfies $r_2\ge r_{\rm ex}$, the DEC is respected on and outside the event horizon.
Thus, violation of the NEC, WEC, or DEC is always hidden inside the event horizons.
On the other hand, by Eqs.~(\ref{NEC}) and (\ref{SEC}), the SEC is respected in the region $r^2\ge r_{\rm SEC}^2(\theta)$, where
\begin{align}
r_{\rm SEC}(\theta):=\frac{l}{\sqrt{2}}\left(1+\sqrt{1+\frac{8a^2}{l^2}\cos^2\theta}\right)^{1/2}(>l).
\end{align}

\subsubsection{Geodesic equations and photon spheres}

Closing this section, we examine whether our non-singular black holes can be distinguished from the Schwarzschild or Kerr black hole by astrophysical observations.
For this purpose, we first derive geodesic equations in the G\"urses-G\"ursey spacetime~(\ref{BL}).

Let $\gamma$ be a geodesic affinely parametrized as $x^\mu=(t(\lambda),r(\lambda),\theta(\lambda),\phi(\lambda))$ in the spacetime~(\ref{BL}), where $\lambda$ is an affine parameter. 
Let $k^\mu=({\dot t}, {\dot r}, {\dot \theta},{\dot \phi})$ be the tangent vector of $\gamma$, where a dot denotes differentiation with respect to $\lambda$.
The components of $k^\mu$ satisfy
\begin{align}
\varepsilon=&-\biggl(1-\frac{2M(r)r}{\Sigma}\biggl){\dot t}^2-\frac{4aM(r)r\sin^2\theta}{\Sigma}{\dot t}{\dot \phi} \nonumber \\
&+\frac{\Sigma}{\Delta}{\dot r}^2+\Sigma{\dot \theta}^2+\biggl(r^2+a^2+\frac{2a^2M(r)r\sin^2\theta}{\Sigma}\biggl)\sin^2\theta{\dot \phi}^2,\label{ds-gamma}
\end{align}
where $\varepsilon=-1,0,1$ correspond to $\gamma$ being timelike, null, and spacelike, respectively.
The spacetime (\ref{BL}) admits Killing vectors $\xi^\mu(\partial/\partial x^\mu)=\partial/\partial t$ and $\Phi^\mu(\partial/\partial x^\mu)=\partial/\partial \phi$, so that $E:=-\xi_\mu k^\mu$ and $L:=\Phi_\mu k^\mu$ are constant along $\gamma$.
$E$ and $L$ represent the energy and angular momentum of the geodesic particle, respectively.
These conservation equations are written as
\begin{align}
E=&\biggl(1-\frac{2M(r)r}{\Sigma}\biggl){\dot t}+\frac{2aM(r)r\sin^2\theta}{\Sigma}{\dot \phi},\label{geodesic-E}\\
L=&-\frac{2aM(r)r\sin^2\theta}{\Sigma}{\dot t}+\biggl(r^2+a^2+\frac{2a^2M(r)r\sin^2\theta}{\Sigma}\biggl)\sin^2\theta{\dot \phi}.\label{geodesic-L}
\end{align}
In addition, the spacetime (\ref{BL}) admits the following two-rank Killing tensor:
\begin{align}
K^{\mu\nu}=\Sigma(\eta^\mu \zeta^\nu+\zeta^\mu \eta^\nu )+r^2 g^{\mu\nu},
\end{align}
which satisfies $\nabla_{(\rho}K_{\mu\nu)}=0$, where $\eta^\mu$ and $\zeta^\mu$ are null vectors defined by 
\begin{align}
\eta^\mu=\biggl(\frac{r^2+a^2}{\Delta},1,0,\frac{a}{\Delta}\biggl),\qquad \zeta^\mu=\biggl(\frac{r^2+a^2}{2\Sigma},-\frac{\Delta}{2\Sigma},0,\frac{a}{2\Sigma}\biggl).
\end{align}
Non-zero components of $K_{\mu\nu}$ are given by
\begin{align}
&K_{tt}=\frac{a^2(\Sigma-2M(r)r\cos^2\theta)}{\Sigma},\\
&K_{t\phi}(=K_{\phi t})=-\frac{a\sin^2\theta(r^4+a^2r^2+\Delta a^2\cos^2\theta)}{\Sigma},\\
&K_{rr}=-\frac{\Sigma a^2\cos^2\theta}{\Delta},\qquad K_{\theta\theta}=r^2\Sigma,\\
&K_{\phi\phi}=\frac{\sin^2\theta}{\Sigma}\left\{\Delta a^4\sin^2\theta\cos^2\theta+ r^2(r^2+a^2)^2\right\}.
\end{align}
As a consequence, $C:=K_{\mu\nu}k^\mu k^\nu$ is constant along $\gamma$.
Then, combining it with Eqs.~(\ref{ds-gamma}), (\ref{geodesic-E}), and (\ref{geodesic-L}), we finally obtain the following set of ordinary differential equations to determine $x^\mu(\lambda)$ along $\gamma$:
\begin{align}
{\dot t}=&\frac{Ea^2\cos^2\theta\Delta + 2aM(r)r(aE - L) + Er^2(r^2+a^2)}{\Delta\Sigma},\label{g-master-t}\\
{\dot r}^2=&\frac{(\varepsilon r^2-C)\Delta+[(r^2+a^2)E-aL]^2}{{\Sigma}^2},\label{g-master-x}\\
{\dot\theta}^2=&\frac{-(E^2 + \varepsilon)a^2\sin^4\theta+( \varepsilon a^2+C+ 2aEL)\sin^2\theta-L^2}{\Sigma^2\sin^2\theta},\label{g-master-theta}\\
{\dot\phi}=&\frac{-2M(r) r(L- Ea\sin^2\theta)+ L\Sigma}{\Delta\Sigma\sin^2\theta}.\label{g-master-phi}
\end{align}
The numerators in the right-hand sides of Eqs.~(\ref{g-master-x}) and (\ref{g-master-theta}) must be non-negative.
In the spherically symmetric case ($a=0$), Eqs.~(\ref{g-master-t})--(\ref{g-master-phi}) reduce to
\begin{align}
\label{g-master2}
\begin{aligned}
{\dot t}=&\frac{Er^2}{r^2-2M(r)r},\qquad {\dot r}^2+V_{\rm eff}(r)=E^2,\\
{\dot\theta}^2=&\frac{C\sin^2\theta-L^2}{r^4\sin^2\theta},\qquad {\dot\phi}=\frac{L}{r^2\sin^2\theta},
\end{aligned}
\end{align}
where the effective potential $V_{\rm eff}(r)$ is defined by 
\begin{align}
V_{\rm eff}(r):=\frac{(C-\varepsilon r^2)(r^2-2M(r)r)}{r^4}.\label{def-V}
\end{align}
For geodesics with constant $\theta(=\theta_0$), $C=L^2/\sin^2\theta_0$ is satisfied.

\begin{figure}[htbp]
\begin{center}
%\rotatebox{-90}{
\includegraphics[width=0.55\linewidth]{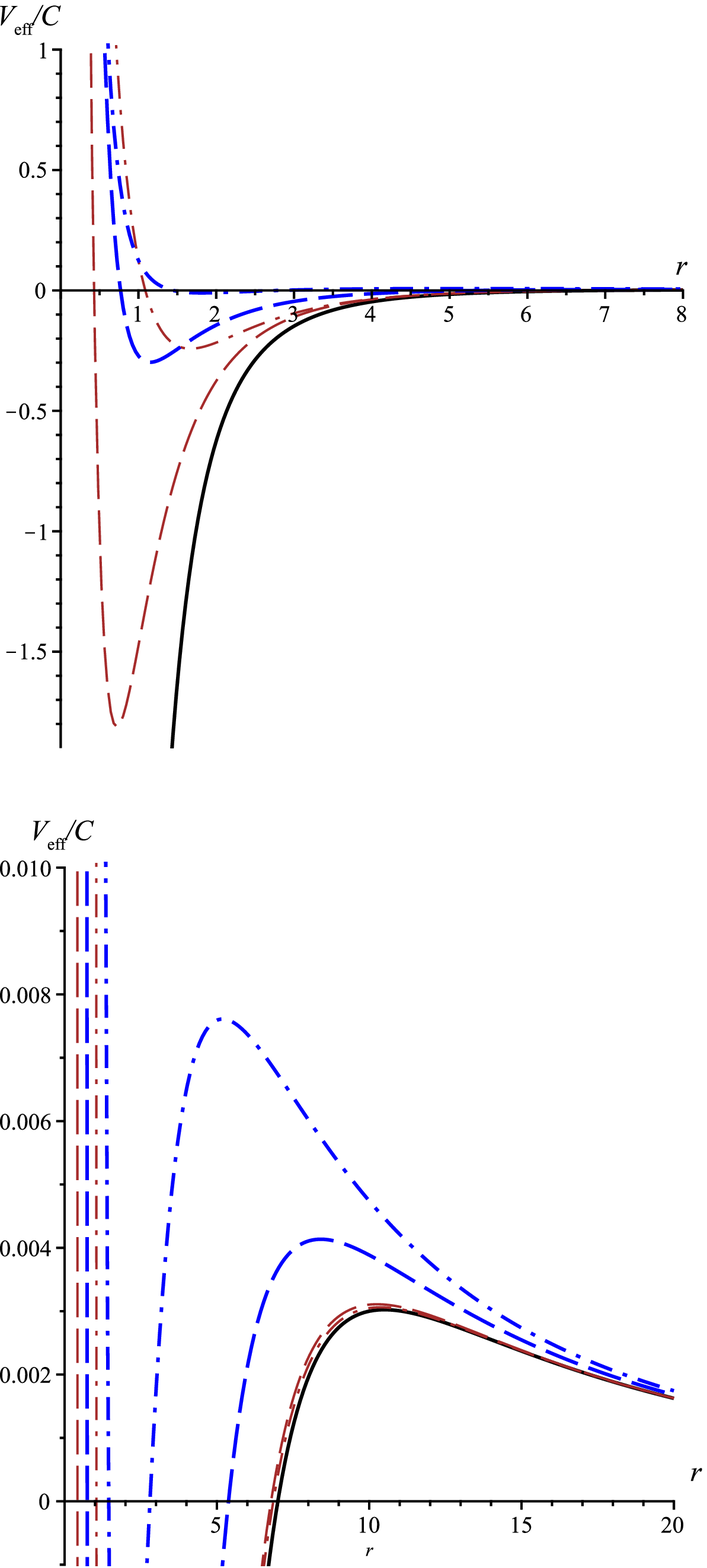}
%\subfigure[]{\includegraphics[width=0.7\linewidth]{Roberts-lambda1.eps}}
%\subfigure[]{\includegraphics[width=0.7\linewidth]{Roberts-lambda2.eps}}
%}
\caption{\label{Veff-m=7over2} The effective potential (\ref{Veff-null}) for null geodesics with $C>0$ for the Schwarzschild (thick solid), Bardeen (thin dashed), Hayward (thin dash-dotted), Dymnikova (thick dashed), and Fan-Wang black holes (thick dash-dotted) with $m=3.5$ and $l=1$.
}
\end{center}
\end{figure}

Hereafter we focus on null geodesics ($\varepsilon=0$) with $C>0$, for which the effective potential becomes
\begin{align}
V_{\rm eff}(r)=\frac{C}{r^2}\biggl(1-\frac{2M(r)}{r}\biggl). \label{Veff-null}
\end{align}
The typical shapes of $V_{\rm eff}(r)$ for null geodesics with $C>0$ for our non-singular black holes are shown in Fig.~\ref{Veff-m=7over2}.
It is observed that $V_{\rm eff}(r)$ has a local maximum at $r=r_{\rm p}$ outside the event horizon, which is the radius of a photon sphere.
With a fixed value of $l$, the value of $r_{\rm p}$ increases as $m$ increases.
It is also observed that the shapes of $V_{\rm eff}(r)$ for the Bardeen and Hayward black holes are similar to the Dymnikova and Fan-Wang black holes, which suggests that the violations of the energy conditions do not qualitatively change the geodesic behaviors.
Actually, these shapes of $V_{\rm eff}(r)$ are similar to the one for the Reissner-Nordstr\"om black hole.
We emphasize that, while the Reissner-Nordstr\"om black hole does not exist in nature because it is easily neutralized by attracting charged particles with an opposite sign, our non-singular black holes are not ``neutralized'' since $l$ is a parameter in the action characterizing the theory that admits such a black hole as a solution.
%------------<fig>---------------------------
%--------------<fig>-----------------------

Now we estimate whether our non-singular black holes can be distinguished from the Schwarzschild black hole when the mass and the size of a photon sphere of a black hole are measured observationally.
By the following expression
\begin{align}
V_{\rm eff}'=-\frac{2C(rM'-3M+r)}{r^4},
\end{align}
the radius of a photon sphere is determined by an algebraic equation $r_{\rm p}=3M(r_{\rm p})-r_{\rm p}M'(r_{\rm p})$.
Since $M'>0$ holds for our non-singular black holes (\ref{m-B})--(\ref{m-FW}) with $m>0$, we obtain $r_{\rm p}<3M(r_{\rm p})$.
For comparison, $r_{\rm p}=3m$ holds for the Schwarzschild black hole with $M(r)=m$.
Here we assume that $M(r)$ obeys $M(r)\simeq m(1-q l/r)$ as $r\to \infty$, where $q$ is a dimensionless constant of order unity.
For the Dymnikova and Fan-Wang black holes, we have $q=4/\pi$ and $q=3$ by Eqs.~(\ref{m-D-r=infty}) and (\ref{m-FW-r=infty}), respectively.
Then, for a large photon sphere, $r_{\rm p}$ is determined by $r_{\rm p}^2-3mr_{\rm p}+4q lm\simeq 0$, which is solved to give
\begin{align}
r_{\rm p}\simeq\frac{3m}{2}\biggl(1+\sqrt{1-\frac{16q l}{9m}}\biggl).\label{rp-large}
\end{align}
For $m\gg l$, we obtain
\begin{align}
\frac{r_{\rm p}}{3m}\simeq 1-\frac{4q l}{9m}+{\cal O}((l/m)^2),\label{rp-large2}
\end{align}
which shows that the difference of the radius $r_{\rm p}$ for a large non-singular black hole from $3m$ for the Schwarzschild black hole is of the order of $l/m(=lc^2/(Gm))$.
Since the length parameter $l$ in the action of the theory should be typically the Planck length $l_{\rm pl}\simeq 1.616229\times 10^{-35}{\rm m}$, we write $l$ as $l=\sigma l_{\rm pl}$, where $\sigma$ is a dimensionless constant.
Then, for a black hole with the solar mass $m=M_\odot\simeq 1.9884\times 10^{30}{\rm kg}$, we have $l/m=\sigma m_{\rm pl}/M_\odot\simeq 1.0946\sigma\times 10^{-38}$, where $m_{\rm pl}\simeq 2.176470\times 10^{-8}{\rm kg}$ is the Planck mass.
This estimation implies that it is extremely difficult to distinguish our non-singular black holes from the Schwarzschild black hole even if we can measure its mass $m$ and size of the photon sphere $r_{\rm p}$ by observations.

%======================================%
%<<<<<<<<<<<< SECTION 1 >>>>>>>>>>>>>>%
%======================================%
\section{Summary}
\label{sec:summary}

In Sec.~\ref{sec:intro}, we have proposed seven criteria C1--C7 to single out physically reasonable non-singular black-hole models.
According to the purely geometric criteria C1--C5, we have studied spherically symmetric non-singular black-hole spacetimes with a regular center described by the metric (\ref{metric-app}) and their rotating counterparts described by the G\"urses-G\"ursey metric (\ref{BL}) without specifying a gravitation theory.
In Sec.~\ref{sec:metric}, we have clarified the relations between the standard energy conditions and asymptotic behaviors of the mass function $M(r)$ around a regular center and an asymptotically flat region satisfying in the spherically symmetric case.
We have also shown that, if a spherically symmetric spacetime (\ref{metric-app}) admits a regular center, its rotating counterpart contains a ring-like conical singularity at $(r,\theta)=(0,\pi/2)$.
Hence, the criterion C1 cannot be satisfied in this class of rotating non-singular black holes.
Although this conical singularity at $(r,\theta)=(0,\pi/2)$ is not a scalar polynomial curvature singularity, it remains an open question whether it is a p.p. curvature singularity or not.

In Sec.~\ref{sec:RC}, we have investigated four particular spherically symmetric non-singular black-hole spacetimes and their rotating counterparts, of which results are summarized in Tables~\ref{table:summary} and \ref{table:summary-rotating}.
Among these four, only the Hayward spacetime respects the LCC and hence the criterion C5.
Since the effective energy-momentum tensors ${\tilde T}_{\mu\nu}(:=G_{\mu\nu})$ for the well-known Bardeen and Hayward black-hole spacetimes do not satisfy the DEC in an asymptotically flat region as shown in Table~\ref{table:BH}, these spacetimes are discarded according to the criterion C3.
In contrast, the Dymnikova and Fan-Wang black holes respect the DEC everywhere.
Although this type of non-singular black holes violate the SEC around the regular center $r=0$ in general, it simply means that the gravitational force becomes repulsive there.

The rotating Dymnikova and Fan-Wang spacetimes can be extended beyond the $r=0$ surface inside the ring-like conical singularity into the region with negative $r$.
We have shown that the rotating Fan-Wang black-hole spacetime contains a curvature singularity in the negative region of $r$.
In contrast, the rotating Dymnikova black-hole spacetime is free from curvature singularities except at $(r,\theta)=(0,\pi/2)$ and also from closed timelike curves.
This fascinating property stems from the fact that the Dymnikova spacetime is invariant for $r\to -r$ and $a\to -a$.
In addition, we have shown that, although all the standard energy conditions are violated around $r=0$ in the rotating Dymnikova black-hole spacetime, the DEC is respected on and outside the event horizons.
As a result, the Dymnikova black-hole spacetime satisfies the criteria C1--C4 and its rotating counterpart satisfies C2--C4. 
Although the rotating counterpart does not satisfy the criterion C1, it satisfies the weaker criterion W-C1 if $(r,\theta)=(0,\pi/2)$ is not a p.p. curvature singularity.
Hence, among the four non-singular black-hole spacetimes with a regular center considered in the present paper, the Dymnikova black-hole spacetime is a preferable model.

Needless to say, an important task is to identify the theory which admits the Dymnikova black hole as a solution without a fine-tuning of the integration constants to satisfy the criterion C6.
In the two-dimensional case, the most general dilatonic action to give the second-order field equations is known and one can construct the action from a given metric of a non-singular black hole~\cite{Kunstatter:2015vxa,Takahashi:2018yzc}.
In fact, this most general two-dimensional dilatonic action can be obtained by a dimensional reduction imposing spherical symmetry from an $n(\ge 4)$-dimensional action made of non-polynomial curvature invariants~\cite{Colleaux:2017ibe,Colleaux:2019ckh}.
On the other hand, it was shown that the Hayward black hole can be a solution in Degenerate Higher-Order Scalar-Tensor (DHOST) theories~\cite{Babichev:2020qpr}, which is a generalization of the most general scalar-tensor theory with the second-order field equation, the Horndeski theory, not to suffer from the Ostrogradsky ghost instability.
Since the Lagrangian of DHOST theories contains several arbitrary functions of the scalar field $\phi$ and the kinetic term $(\nabla\phi)^2$, a class of such theories could allow the Dymnikova black hole as a generic configuration.

Once such a theory is identified, the subsequent task is to study the dynamical stability of the Dymnikova black hole according to the criterion C7.
In this context, in addition to the stability of the region outside the event horizon, stability of an inner horizon is also non-trivial.
While it is well-known that the inner horizon of the Reissner-Nordstr\"om black hole suffers from the mass-inflation instability~\cite{Poisson:1989zz,Poisson:1990eh}, two different results have been obtained in the case of spherically symmetric non-singular black holes with a regular center~\cite{Bonanno:2020fgp,Rubio:2021obb}. 
Because these works were performed in the framework of general relativity, the results cannot be directly applied to the black holes in generalized theories of gravity.
For example, it has been reported in Brans-Dicke theory that mass-inflation occurs in accreting black holes~\cite{Avelino:2009vv}. 
In contrast, it has been reported in Eddington-inspired Born-Infeld gravity that, different from general relativity, there is a minimum accretion rate below which mass inflation does not occur~\cite{Avelino:2015fve,Avelino:2016kkj}.
From this point of view, the Dymnikova black-hole spacetime should be modified to satisfy the LCC because the problem of mass inflation can be effectively solved then.
These important tasks are left for future investigations.

\subsection*{Acknowledgements}
The author thanks Tomohiro Harada, Gabor Kunstatter, and Jos\'{e} M. M. Senovilla for helpful comments.

\appendix

%======================================%
%<<<<<<<<<<<< SECTION 1 >>>>>>>>>>>>>>%
%======================================%
\section{Birkhoff prohibits generic non-singular black holes in GR}
\label{sec:no-go-GR}

In this appendix, we show that the criterion C5 in Sec.~\ref{sec:intro} cannot be respected in general relativity with a matter field which leads to Birkhoff's theorem.
A well-known class of nonlinear electromagnetic fields are examples of such a matter field.

\subsection{Generalized Birkhoff's theorem in general relativity}

We consider the following most general $n(\ge 4)$-dimensional spherically symmetric spacetime;
\begin{align}
\D s^2=g_{AB}(y)\D y^A\D y^B +r^2(y)\gamma_{ij}(z)\D z^i\D z^j,
\label{eq:ansatz}
\end{align} 
where $A,B=0,1$ and $i,j=2,3,\cdots,n-1$.
$g_{AB}$ is a metric on a two-dimensional Lorentzian spacetime $M^2$ and $r(y)$ is a scalar on $M^2$.
$\gamma_{ij}$ is a metric on an $(n-2)$-dimensional unit sphere $S^{n-2}$ satisfying ${}^{(n-2)}R_{ijkl}=\gamma_{ik}\gamma_{jl}-\gamma_{il}\gamma_{jk}$, where ${}^{(n-2)}R_{ijkl}$ is the Ricci tensor constructed from $\gamma_{ij}$.
The metric (\ref{eq:ansatz}) gives the Einstein tensor in the following form~\cite{Maeda:2007uu}:
\begin{align}
G_{\mu\nu}\D x^\mu\D x^\nu={\cal G}_{AB}(y)\D y^A\D y^B+{\cal {\bar G}}(y)\gamma_{ij}(z)\D z^i\D z^j,
\label{eq:ansatz-G}
\end{align} 
where 
\begin{align}
{\cal G}_{AB}:=&-(n-2)\frac{D_AD_Br}{r}+g_{AB}\biggl\{\frac{(n-2)D^2r}{r}-\frac{(n-2)(n-3)[1-(Dr)^2]}{2r^2}\biggl\}, \label{Gab} \\
{\cal {\bar G}}:=&-\frac12r^2({}^{(2)}R)+(n-3)rD^2r-\frac{(n-3)(n-4)}{2}[1-(Dr)^2]. \label{Gij} 
\end{align}
Here ${}^{(2)}R$ is the Ricci scalar of $M^2$, $D_A$ is the covariant derivative on $M^2$, and we have defined $(Dr)^2:=g^{AB}(D_Ar)(D_Br)$ and $D^2r:=D^AD_Ar$.
For compatibility with the Einstein equations, the energy-momentum tensor $T_{\mu\nu}$ has to be in the following form:
\begin{align}
T_{\mu\nu}\D x^\mu\D x^\nu={T}_{AB}(y)\D y^A\D y^B+r(y)^2p_2(y)\gamma_{ij}(z)\D z^i\D z^j,\label{eq:ansatz-T}
\end{align} 
where ${T}_{AB}$ and $p_2$ are a two-tensor and a scalar on $M^2$, respectively.

The following Birkhoff's theorem was shown in~\cite{Dymnikova:1992ux} with $n=4$ and a particular case of the more general result in~\cite{Bronnikov:1994ja,Das:2001md} with arbitrary $n(\ge 4)$.
%----------------------- lemma ------------------------------%
\begin{Prop}
\label{Prop:Birkhoff}
Suppose that $(D r)^2\ne 0$ holds and $T_{AB}$ in Eq.~(\ref{eq:ansatz-T}) satisfies $T^A_{~B}=-\rho(y)\delta^A_{~B}$ in the $n(\ge 4)$-dimensional spacetime (\ref{eq:ansatz}).
Then, the general solution of the Einstein equations is given by 
\begin{align}
\D s^2=&-f(r)\D t^2+f(r)^{-1}\D r^2+r^2\gamma_{ij}\D z^i\D z^j,\label{metric-Birkhoff}\\
&\rho=\rho(r), \quad p_2=p_2(r),
\end{align} 
where $f(r)$ and $p_2(r)$ are determined by $\rho(r)$ as
\begin{align}
&f(r)=1-\frac{2{\bar M}}{r^{n-3}}-\frac{2 }{(n-2)r^{n-3}}\int^rx^{n-2}\rho(x)\D x,\label{f}\\
&p_2(r)=-\frac{(n-2)\rho+r\rho'}{n-2}.\label{fluid-eq}
\end{align}
Here ${\bar M}$ is an integration constant and the integral in the last term of Eq.~(\ref{f}) does not generate an additional integration constant.
\end{Prop}
{\it Proof:}
If $(D r)^2\ne 0$ holds in the spacetime (\ref{eq:ansatz}), we can use $r$ as a coordinate on $M^2$ without loss of generality such that $(y^0,y^1)=(t,r)$ and then the metric can be expressed as 
\begin{align}
\D s^2=&-f(t,r)e^{-2\delta(t,r)}\D t^2+f(t,r)^{-1}\D r^2+r^2\gamma_{ij}\D z^i\D z^j.\label{metric}
\end{align} 
By assumption, the energy-momentum tensor (\ref{eq:ansatz-T}) is now in the following form:
\begin{align}
T^\mu_{~\nu}=\mbox{\rm diag}(-\rho,-\rho,p_2,\cdots,p_2). \label{matter-Birkhoff}
\end{align} 
By the following expressions
\begin{align}
G^{t}_{~r}=\frac{(n-2)f_{,t}}{2rf^2}e^{2\delta}, \qquad G^{r}_{~t}=-\frac{(n-2)f_{,t}}{2r},\label{lm:ff1}
\end{align} 
where a comma denotes partial differentiation, the components $G^{t}_{~r}=0$ and $G^{r}_{~t}=0$ of the Einstein equations show $f(t,r)=f(r)$.
From the Einstein equations $G^{t}_{~t}-G^{r}_{~r}=T^{t}_{~t}-T^{r}_{~r}=0$ with the expression $G^{t}_{~t}-G^{r}_{~r}=(n-2)r^{-1}\delta'f$, we obtain $\delta=\delta(t)$.
We can set $\delta(t)=0$ without loss of generality by a redefinition of $t$ such that $e^{-\delta(t)}\D t\to \D t$, and then the metric reduces to the form (\ref{metric-Birkhoff}).
The Einstein equations with the metric (\ref{metric-Birkhoff}) require $\rho=\rho(r)$ and $p_2=p_2(r)$.
The component $G^{t}_{~t}=-\rho$ (or equivalently $G^{r}_{~r}=-\rho$) is integrated to give 
\begin{align}
f(r)=&\frac{1}{(n-2)r^{n-3}}\int^r\left\{(n-2)(n-3)-2 x^2\rho(x)\right\}x^{n-4}\D x \nonumber \\
=&1-\frac{2{\bar M}}{r^{n-3}}-\frac{2 }{(n-2)r^{n-3}}\int^rx^{n-2}\rho(x)\D x.\label{f-ret}
\end{align}
Substituting the metric function (\ref{f-ret}) into the component $G^{i}_{~j}=p_2\delta^i_{~j}$, we obtain the expression (\ref{fluid-eq}).
\qed
%----------------------- lemma ------------------------------%

\noindent
As a corollary of Proposition~\ref{Prop:Birkhoff}, a regular center is shown to be non-generic, which is a generalization of the claim in~\cite{Chinaglia:2017uqd} for $n=4$.
%----------------------- lemma ------------------------------%
\begin{Coro}
\label{Coro:nogo}
Under the assumptions in Proposition~\ref{Prop:Birkhoff}, $r=0$ is generically singular.
\end{Coro}
{\it Proof:}
By Proposition~\ref{Prop:Birkhoff}, the system reduces to Eqs.~(\ref{metric-Birkhoff})--(\ref{fluid-eq}).
Then, for ${\bar M}\ne 0$, the metric function $f(r)$ given by Eq.~(\ref{f}) blows up as $r\to 0$ for any $\rho(x)$, so that $r=0$ is a scalar curvature singularity without a fine-tuning ${\bar M}=0$.
\qed
%----------------------- lemma ------------------------------%

\noindent
A regular center requires ${\bar M}=0$ and $\rho(r)\simeq \rho_0r^{\chi}+{O}(r^{\chi+\epsilon})$ around $r=0$ with $\chi\ge 0$, $\epsilon>0$, and $\rho_0\ne 0$, with which Eq.~(\ref{f}) gives
\begin{align}
\lim_{r\to 0}f(r)\simeq 1-\frac{2\rho_0}{(n-2)(n-1+\chi)}r^{2+\chi}+{O}(r^{2+\chi+\epsilon}).
\end{align}

\subsection{A class of nonlinear electromagnetic fields}
\label{App:NEF}

Here we show that Proposition~\ref{Prop:Birkhoff} and Corollary~\ref{Coro:nogo} can be applied to a class of nonlinear electromagnetic fields, of which action is given by 
\begin{align}
S=\int\D^n x\sqrt{-g}\biggl(\frac{1}{2}R-\beta {\cal L}(X)\biggl),\label{action-FW-n}
\end{align} 
where $X:=\alpha F_{\rho\sigma}F^{\rho\sigma}$ and the Faraday tensor $F_{\mu\nu}$ is given in terms of a gauge field $A_\mu$ as $F_{\mu\nu} :=\partial_\mu A_\nu-\partial_\nu A_\mu$.
We have introduced a constant $\alpha$ to make $X$ be dimensionless and assume that ${\cal L}(X)$ is not a constant function.
In the case of a standard Maxwell field, we have ${\cal L}(X)=X$, so that a combination $\alpha\beta$ represents a single coupling constant.
Varying the action (\ref{action-FW-n}), we obtain the Einstein equations $G_{\mu\nu}=T_{\mu\nu}$ with
\begin{align}
T_{\mu\nu}=&4\beta\biggl(\alpha {\cal L}_{,X}F_{\mu\rho}F_{\nu}^{~\rho}-\frac14g_{\mu\nu} {\cal L}\biggl) \label{FW-em}
\end{align} 
and the following field equations 
\begin{align}
\nabla_\nu({\cal L}_{,X}F^{\mu\nu})=0.\label{max}
\end{align} 
The Bianchi identity $\nabla_{[\rho}F_{\mu\nu]}=0$ is automatically satisfied for $F_{\mu\nu} :=\partial_\mu A_\nu-\partial_\nu A_\mu$.

Under an assumption, Proposition~\ref{Prop:Birkhoff} can be applied to the system (\ref{action-FW-n}) and therefore a regular center is non-generic by Corollary~\ref{Coro:nogo}.
%----------------------- lemma ------------------------------%
\begin{Prop}
\label{Prop:nogo-Max}
Suppose that a gauge field $A_\mu$ has the following form
\begin{align} 
A_\mu\D x^\mu=A_B(y)\D y^B+A_i(z)\D z^i \label{A-anzats}
\end{align}
in an $n(\ge 4)$-dimensional spacetime (\ref{eq:ansatz}) in the system (\ref{action-FW-n}), so that $F_{\mu\nu}$ is given by 
\begin{align} 
F_{\mu\nu}\D x^\mu\wedge \D x^\nu=F_{AB}(y)\D y^A\wedge \D y^B+F_{ij}(z)\D z^i\wedge \D z^j,\label{F-anzats}
\end{align}
where $F_{AB}=\partial_AA_B-\partial_BA_A$ and $F_{ij}=\partial_iA_j-\partial_jA_i$.
Then, if $(Dr)^2\ne 0$ holds, the system reduces to Eqs.~(\ref{metric-Birkhoff})--(\ref{fluid-eq}) in Proposition~\ref{Prop:Birkhoff} with 
\begin{align} 
&X=\alpha\left\{-2{F_{tr}}^2+(n-2)Q_m^2r^{-4}\right\},\label{X2}\\
&\rho(=-p_1)=4\beta\biggl(\alpha {\cal L}_{,X}{F_{tr}}^2+\frac14 {\cal L}\biggl),\label{NLE-rho}\\
&p_2=4\beta\biggl(\alpha r^{-4}{\cal L}_{,X}Q_m^2-\frac14 {\cal L}\biggl).\label{NLE-p2}
\end{align} 
Here $F_{tr}$ and $F_{ij}$ are determined by the following equations
\begin{align}
&{\cal L}_{,X}F_{tr}=\frac{Q_e}{r^{n-2}},\label{cons-NEM}\\
&{\cal L}_{,X}{\hat D}_j F^{ij}=0, \label{harmonic-z2}\\
&\gamma^{kl}F_{ik}F_{jl}=Q_m^2\gamma_{ij},\label{mag-F}
\end{align}
where $Q_e$ and $Q_m$ are constants and ${\hat D}_i$ is the covariant derivative on $S^{n-2}$.
\end{Prop}
{\it Proof:}
From the expression (\ref{F-anzats}), we obtain 
\begin{align}
T_{ij}=&4\beta\biggl(\alpha r^{-2}{\cal L}_{,X}\gamma^{kl}F_{ik}F_{jl}-\frac14r^2\gamma_{ij} {\cal L}\biggl).\label{Tij}
\end{align} 
For compatibility with Eq.~(\ref{eq:ansatz-G}) through the Einstein equations $G_{\mu\nu}=T_{\mu\nu}$, the magnetic components $F_{ij}$ must satisfy Eq.~(\ref{mag-F}).
The Bianchi identity $\nabla_{[\rho} F_{\mu\nu]}=0$ gives
\begin{align} 
D_{[C} F_{AB]}=0,\qquad {\hat D}_{[k} F_{ij]}=0.\label{Bianchi-Max}
\end{align}

Since $F_{AB}$ is anti-symmetric, $F^{AD}F_{BD}=F_{01}F^{01}\delta^A_{~B}$ holds and hence $X(:=\alpha F_{\rho\sigma}F^{\rho\sigma})$ and non-zero components of the energy-momentum tensor (\ref{FW-em}) are given by 
\begin{align} 
X=&\alpha\left\{2F_{01}F^{01}+(n-2)Q_m^2r^{-4}\right\},\\
T^A_{~B}=&4\beta\biggl(\alpha {\cal L}_{,X}F_{01}F^{01}-\frac14 {\cal L}\biggl)\delta^A_{~B},\label{Tab2}\\
T^i_{~j}=&4\beta\biggl(\alpha r^{-4}{\cal L}_{,X}Q_m^2-\frac14 {\cal L}\biggl)\delta^i_{~j}.\label{Tij2}
\end{align} 
Since $T_{\mu\nu}$ is given in the form (\ref{eq:ansatz-T}) and $T^A_{~B}\propto \delta^A_{~B}$ holds, the system reduces to Eqs.~(\ref{metric-Birkhoff})--(\ref{fluid-eq}) with Eqs.~(\ref{X2})--(\ref{NLE-p2}) by Proposition~\ref{Prop:Birkhoff}.

Because of ${\cal L}_{,X}={\cal L}_{,X}(y)$ due to Eq.~(\ref{X2}), the field equations (\ref{max}) with $\mu=A$ give
\begin{align} 
D_B(r^{n-2}{\cal L}_{,X}F^{AB})=0~~\leftrightarrow~~\frac{\partial}{\partial y^B}\left(r^{n-2}{\cal L}_{,X}F^{AB}\right)=0.\label{max-M2}
\end{align} 
In the spacetime (\ref{metric-Birkhoff}), the above equations are integrated to give Eq.~(\ref{cons-NEM}).
On the other hand, the field equations (\ref{max}) with $\mu=i$ become Eq.~(\ref{harmonic-z2}) as
\begin{align} 
0=\nabla_\nu({\cal L}_{,X}F^{i\nu})=\frac{{\cal L}_{,X}}{\sqrt{\gamma}}\frac{\partial}{\partial z^j}\left(\sqrt{\gamma}F^{ij}\right)={\cal L}_{,X}{\hat D}_j F^{ij},
\end{align} 
where $\gamma:=\det(\gamma_{ij})$. 
\qed
%----------------------- lemma ------------------------------%

We note that $Q_m=0$ and $F_{ij}\equiv 0$ are equivalent since Eq.~(\ref{mag-F}) gives $F_{ij}F^{ij}=(n-2)Q_m^2$ and $\gamma_{ij}$ is an Euclidean metric.
If ${\cal L}_{,X}\ne 0$ is satisfied, Eq.~(\ref{harmonic-z2}) gives
\begin{align}
{\hat D}_j F^{ij}=0, \label{harmonic-z}
\end{align}
which means that $F^{ij}$ is a harmonic two-form on $S^{n-2}$.
As a consequence, nontrivial configurations of $F_{ij}$, namely magnetic solutions, are not allowed in higher dimensions ($n\ge 5$), which is shown by a corollary of Hodge's theorem~\cite{Maeda:2010qz}.
(See Theorem 7.8 and Eq.~(7.198) in the textbook~\cite{Nakahara} for the proof of Hodge's theorem.)
%----------------------- lemma ------------------------------%
\begin{The}
\label{The:p-form}
({\it Hodge's theorem.})
The $p(\ge 0)$-th Betti number $b_p(M)$ of a compact Riemannian manifold $M$ is equal to the dimension of the space of harmonic $p$-forms on $M$. 
\end{The}
%{\it Proof:}
%See Theorem 7.8 and Eq.~(7.198) in the textbook~\cite{Nakahara}.
%\qed
%----------------------- lemma ------------------------------%
%----------------------- lemma ------------------------------%
\begin{Coro}
\label{Coro:2-form}
${\hat D}_j F^{ij}=0$ on $S^{n-2}$ admits only a trivial solution ($F^{ij}\equiv 0$) if $n\ge 5$.
For $n=4$, the only linearly independent solution is the volume two-form $F_{ij}=\sqrt{\gamma}\epsilon_{ij}$, where $\epsilon_{ij}$ is the Levi-Civit\'a symbol.
\end{Coro}
{\it Proof:}
Since $S^{n-2}$ is compact and its second Betti number is~\cite{Betti} 
\begin{eqnarray}
b_2(S^{n-2}) = \left\{
\begin{array}{ll}
1 & (n=4)\\
0 & (n\ge 5)
\end{array}
\right.,
\end{eqnarray}
by Theorem~\ref{The:p-form}, ${\hat D}_j F^{ij}=0$ admits only one linearly independent solution for $n=4$ and only a trivial solution  for $n\ge 5$.
One can check that $F_{ij}=\sqrt{\gamma}\epsilon_{ij}$ solves ${\hat D}_j F^{ij}=0$ for $n=4$ by direct calculations.
\qed
%----------------------- lemma ------------------------------%

Eventually, if ${\cal L}_{,X}\ne 0$ holds in the system (\ref{action-FW-n}), the general solution in Proposition~\ref{Prop:Birkhoff} is expressed as follows.
%----------------------- lemma ------------------------------%
\begin{Prop}
\label{Prop:dyonic-sol}
Consider an $n(\ge 4)$-dimensional spherically symmetric spacetime (\ref{eq:ansatz}) satisfying $(Dr)^2\ne 0$ in the system (\ref{action-FW-n}), in which $A_\mu$ is given in the form of Eq.~(\ref{A-anzats}).
Suppose that an algebraic equation ${\cal L}_{,X}(X)=0$ does not admit any real solution.
Then, the general solution for $n\ge 5$ is given by 
\begin{align}
&\D s^2=-f(r)\D t^2+f(r)^{-1}\D r^2+r^2\gamma_{ij}\D z^i\D z^j,\label{metric-dyo}\\
&f(r)=1-\frac{2{\bar M}}{r^{n-3}}-\frac{8\beta}{(n-2)r^{n-3}}\int^rx^{n-2}\left(\alpha{\cal L}_{,X}{F_{tr}}^2+\frac14{\cal L}\right)\D x,\label{dyo-sol1}\\
&F_{\mu\nu} \D x^\mu\wedge\D x^\nu=2F_{tr}\D t\wedge \D r,\label{dyo-sol2}
\end{align}
where ${\bar M}$ is a constant and $F_{tr}(r)$ is determined algebraically by 
\begin{align}
{\cal L}_{,X}F_{tr}=\frac{Q_e}{r^{n-2}},\qquad X=-2\alpha{F_{tr}}^2 \label{Ftr-det}
\end{align}
with an arbitrary constant $Q_e$.
The general solution for $n=4$ is given 
\begin{align}
&\D s^2=-f(r)\D t^2+f(r)^{-1}\D r^2+r^2(\D\theta^2+\sin^2\theta\D\phi^2),\label{metric-mag}\\
&F_{\mu\nu} \D x^\mu\wedge\D x^\nu=2F_{tr}\D t\wedge \D r+2Q_m\sin\theta\D\theta\wedge\D\phi,\label{mag-sol2}
\end{align}
where $f(r)$ is given by Eq.~(\ref{dyo-sol1}) with $n=4$ and $F_{tr}(r)$ is determined algebraically by 
\begin{align}
{\cal L}_{,X}F_{tr}=\frac{Q_e}{r^{2}},\qquad X=&-2\alpha\left({F_{tr}}^2-Q_m^2r^{-4}\right) \label{Ftr-det2}
\end{align}
with arbitrary constants $Q_e$ and $Q_m$.
\end{Prop}
{\it Proof:}
By Lemma~\ref{lm:ff}, Proposition~\ref{Prop:Birkhoff} can be applied.
Then, Eq.~(\ref{f}) with Eq.~(\ref{NLE-rho}) gives Eq.~(\ref{dyo-sol1}).
Equations~(\ref{Ftr-det}) and (\ref{Ftr-det2}) to determine $F_{tr}$ are from Eqs.~(\ref{X2}) and (\ref{cons-NEM}).
By Corollary~\ref{Coro:2-form}, $F_{ij}\equiv 0$ holds for $n\ge 5$.
For $n=4$, since $F_{\theta\phi}=\sin\theta$ is the only linearly independent solution of ${\hat D}_j F^{ij}=0$ in the coordinates (\ref{metric-mag}) by Corollary~\ref{Coro:2-form}, its general solution satisfying Eq.~(\ref{mag-F}) is $F_{\theta\phi}=Q_m\sin\theta$.
\qed
%----------------------- lemma ------------------------------%

The general solution (\ref{metric-mag})--(\ref{Ftr-det2}) for $n=4$ was derived in~\cite{Bronnikov:2000vy}.
Proposition~\ref{Prop:dyonic-sol} is a generalization of the Birkhoff's theorem proved in~\cite{Garcia-Diaz:2019acq} for $n=4$. (See also~\cite{Bronnikov:2019fgh} and ~\cite{Garcia-Diaz:2020cpu}.)
In the standard Maxwell case (${\cal L}=X$) for general $n(\ge 4)$, Eqs.~(\ref{NLE-rho}), (\ref{NLE-p2}), (\ref{X2}), and (\ref{cons-NEM}) give
\begin{align} 
&\rho=- p_1=2\alpha\beta\biggl(\frac{Q_e^2}{r^{2(n-2)}}+\frac{(n-2)Q_m^2}{2r^4}\biggl),\\
&p_2=2\alpha\beta\biggl(\frac{Q_e^2}{r^{2(n-2)}}-\frac{(n-6)Q_m^2}{2r^4}\biggl).
\end{align} 
Hence, Eq.~(\ref{f-ret}) gives the following metric function
\begin{align}
f(r)=1-\frac{2{\bar M}}{r^{n-3}}+\frac{4\alpha\beta Q_e^2}{(n-2)(n-3)r^{2(n-3)}}-\frac{2\alpha\beta Q_m^2}{(n-5)r^2}.\label{f-gr}
\end{align}
By Corollary~\ref{Coro:2-form}, $Q_m=0$ is required in higher dimensions ($n\ge 5$).

A physically reasonable nonlinear generalization of the Maxwell theory should satisfy the weak-field limit ${\cal L}\simeq X$ (so that ${\cal L}_{,X}\simeq 1$) as $X\to 0$.
However, as shown below, this criterion is not satisfied around the regular center $r=0$ in the spherically symmetric solution with $Q_e\ne 0$ in Proposition~\ref{Prop:dyonic-sol}.
This is a generalization of the result obtained in~\cite{Bronnikov:1979ex} for $n=4$. (See also~\cite{Bronnikov:2000vy}.)

%----------------------- lemma ------------------------------%
\begin{Prop}
\label{Prop:no-go-NED}
Suppose that the spacetime admits a regular center $r=0$ in the general spherically symmetric solution in Proposition~\ref{Prop:dyonic-sol}.
Then, $Q_e\ne 0$ implies (i) $Q_m=0$ and (ii) $|{\cal L}_{,X}|\to \infty$ and $X\to 0$ hold as $r\to 0$.
\end{Prop}
{\it Proof:}
Since the metric $g_{\mu\nu}$ and its inverse $g^{\mu\nu}$ as well as the Riemann tensor $R^{\mu\nu}_{~~\rho\sigma}$ are finite in the limit to a regular center $r\to 0$, $G^\mu_{~\nu}$ is finite as $r\to 0$.
Then, the Einstein equations show that $\rho$ and $p_2$ given by Eqs.~(\ref{NLE-rho}) and (\ref{NLE-p2}) are finite as well.
Since $\rho+p_2=4\beta({\cal L}_{,X}{F_{tr}}^2+r^{-4}{\cal L}_{,X}Q_m^2)$ is finite as $r\to 0$ and the signs of the two terms in the right-hand side are the same, we obtain
\begin{align} 
&\lim_{r\to 0}|{\cal L}_{,X}{F_{tr}}^2|<\infty,\label{Qe-bound}\\
&\lim_{r\to 0}|r^{-4}{\cal L}_{,X}Q_m^2|<\infty.\label{Qm-bound}
\end{align} 
In the case of $Q_e\ne 0$, Eq.~(\ref{Ftr-det}) gives
\begin{align}
\lim_{r\to 0}{{\cal L}_{,X}}^2{F_{tr}}^2\to \infty.\label{F-bound}
\end{align}
Equations~(\ref{Qe-bound}) and (\ref{F-bound}) show $|{\cal L}_{,X}|\to \infty$ and ${F_{tr}}^2\to 0$ as $r\to 0$, so that Eq.~(\ref{Qm-bound}) requires $Q_m=0$.
Then, by Eqs.~(\ref{Ftr-det}) and (\ref{Ftr-det2}), $X\to 0$ hold as $r\to 0$.
\qed
%----------------------- lemma ------------------------------%

Proposition~\ref{Prop:no-go-NED} shows that a dyonic solution with a regular center is not possible in the four-dimensional system ($n=4$) in Proposition~\ref{Prop:dyonic-sol}.
Accordingly, a non-singular black-hole solution must be purely electric ($Q_m=0$) in $n(\ge 4)$ dimensions or purely magnetic ($Q_e=0$) in four dimensions.
However, the proper weak-field limit $\lim_{X\to 0}{\cal L}\simeq X$ is not achieved around the regular center in the former case.

In the purely magnetic solution ($Q_e=0$) in four dimensions, one can easily identify the form of ${\cal L}(X)$ from a given metric function $f(r)$ as demonstrated in~\cite{Fan:2016hvf}.
By Eq.~(\ref{Ftr-det2}) with $Q_e=0$, we can write $r$ as a function of $X$ as
\begin{align} 
r=\biggl(\frac{2\alpha Q_m^2}{X}\biggl)^{1/4}.\label{r-X-mag}
\end{align} 
From Eq.~(\ref{dyo-sol1}) with $n=4$, $F_{tr}=0$, and $M=0$, we obtain
\begin{align}
\beta {\cal L}(X(r))=\frac{1-f(r)-rf'(r)}{r^2}=\frac{2M'(r)}{r^2},
\end{align}
where $M(r)$ is the mass function in Eq.~(\ref{metric-app}).
Replacing $r$ by $X$ using Eq.~(\ref{r-X-mag}), one finds the desired form of $\beta {\cal L}(X)$.

For example, the Dymnikova spacetime with the metric function (\ref{nonsingular-D}) can be a magnetic solution in the four-dimensional ($n=4$) system (\ref{action-FW-n}) with following Lagrangian function:
\begin{align}
{\cal L}(X)=&\frac{X}{(1+|X|^{1/2})^2}.\label{D-BH}
\end{align}
Substituting Eq.~(\ref{Ftr-det2}) with $Q_e=0$ into Eqs.~(\ref{dyo-sol1}) with $n=4$ and $F_{tr}=0$, we obtain the following metric function:
\begin{align}
f(r)=&1-\frac{2{\bar M}}{r}-\frac{\beta q^3}{2 r}\biggl\{\arctan\biggl(\frac{r}{q}\biggl)-\frac{qr}{r^2+q^2}\biggl\}, \label{D-magnetic}\\
q:=&(2\alpha Q_m^2)^{1/4}.
\end{align} 
This metric function with ${\bar M}=0$ is identical to Eq.~(\ref{nonsingular-D}) with $l=q$ and $m=\pi \beta q^3/8$.
Since $\beta$ is not an integration constant but a coupling constant in the action, the solution (\ref{D-magnetic}) contains only one free parameter $q$.
By Eq.~(\ref{Ftr-det2}) with $Q_e=0$, $X\to 0$ is realized only in the asymptotically flat region $r\to \infty$ and Eq.~(\ref{D-BH}) shows that a proper weak-field limit of the Lagrangian function $\lim_{X\to 0}{\cal L}\simeq X$ is achieved there.

Lastly, we should note that ${\cal L}_{,X}\ne 0$ is assumed in Proposition~\ref{Prop:dyonic-sol}.
Actually, depending on the form of the Lagrangian function ${\cal L}(X)$, field equations (\ref{cons-NEM}) and (\ref{harmonic-z2}) may permit an exceptional solution $X=X_{\rm sol}=$constant which satisfies ${\cal L}_{,X}=0$ algebraically.
In such an exceptional solution, Eq.~(\ref{cons-NEM}) shows $Q_e=0$ but $F_{tr}$ may be non-vanishing.
Furthermore, Eq.~(\ref{harmonic-z2}) does not mean that $F_{ij}$ is a harmonic two-form.
Even in such a case, there is no magnetic solution in odd dimensions as shown below.
(The absence of magnetic solution has been shown in a more general class of spacetimes~\cite{Ortaggio:2007hs}.)
%----------------------- lemma ------------------------------%
\begin{lm}
\label{lm:ff}
$F_{ij}\equiv 0$ holds in odd dimensions.
\end{lm}
{\it Proof:}
If $n$ is odd, we obtain $\mbox{det}(F_{ij})= 0$ shown by $\mbox{det}(F_{ij})=\mbox{det}(F_{ji})=\mbox{det}(-F_{ij})=(-1)^{n-2}\mbox{det}(F_{ij})=-\mbox{det}(F_{ij})$.
Taking the determinant of Eq.~(\ref{mag-F}), we obtain $Q_m^2\mbox{det}(\gamma_{ij})=-\mbox{det}(F_{ik})\mbox{det}(\gamma^{kl})\mbox{det}(F_{lj})=0$, which shows $Q_m=0$ (and hence $F_{ij}\equiv 0$) for odd $n$.
\qed
%----------------------- lemma ------------------------------%

As a result, in the system (\ref{action-FW-n}) where ${\cal L}_{,X}(X)=0$ admits a real solution, the general solution in Proposition~\ref{Prop:Birkhoff} is expressed as follows.
%----------------------- lemma ------------------------------%
\begin{Prop}
\label{Prop:dyonic-sol-excep}
Consider the system in Proposition~\ref{Prop:dyonic-sol} and suppose that an algebraic equation ${\cal L}_{,X}(X)=0$ admits a real solution $X=X_{\rm sol}$ such that $F_{tr}$ given by Eq.~(\ref{Ftr-exceptional}) is real.
Then, the general solution consists of the solution in Proposition~\ref{Prop:dyonic-sol} and a solution described by 
\begin{align}
&\D s^2=-f(r)\D t^2+f(r)^{-1}\D r^2+r^2\gamma_{ij}\D z^i\D z^j,\label{metric-dyo-ex}\\
&f(r)=1-\frac{2{\bar M}}{r^{n-3}}-\frac{2\beta{\cal L}(X_{\rm sol})}{(n-1)(n-2)}r^2,\label{dyo-sol1-ex}\\
&F_{\mu\nu} \D x^\mu\wedge\D x^\nu=2F_{tr}(r)\D t\wedge \D r+F_{ij}(z)\D z^i\wedge \D z^j,\label{dyo-sol2-ex}
\end{align}
where $F_{tr}$ is given by 
\begin{align} 
F_{tr}=\pm\sqrt{\frac{(n-2)Q_m^2r^{-4}+X_{\rm sol}}{2}} \label{Ftr-exceptional}
\end{align} 
with $Q_{\rm e}=0$.
In the exceptional solution~(\ref{metric-dyo-ex})--(\ref{dyo-sol2-ex}), $F_{ij}\equiv 0$ holds (hence $Q_m=0$) for odd $n(\ge 5)$ and $F_{ij}(z)$ are determined by Eqs.~(\ref{mag-F}) and (\ref{Bianchi-Max}) for even $n(\ge 4)$.
\end{Prop}
{\it Proof:}
If $X=X_{\rm sol}=$ satisfies ${\cal L}_{,X}=0$ algebraically, Eq.~(\ref{cons-NEM}) gives $Q_e=0$ and $F_{tr}$ becomes Eq.~(\ref{Ftr-exceptional}) by Eq.~(\ref{X2}).
On the other hand, Eq.~(\ref{harmonic-z2}) are trivially satisfied and $F_{ij}\equiv 0$ holds by Lemma~\ref{lm:ff} in odd dimensions.
Hence, in even dimensions, $F_{ij}$ are determined by Eq.~(\ref{mag-F}) and the Bianchi identity~(\ref{Bianchi-Max}) ensures that $F_{ij}$ is given in terms of the potential $A_i(z)$.
\qed
%----------------------- lemma ------------------------------%

\noindent
By Eqs.~(\ref{Tab2}) and (\ref{Tij2}), the energy-momentum tensor for an exceptional solution in Proposition~\ref{Prop:dyonic-sol-excep} is given by $T^\mu_{~\nu}=-\beta{\cal L}(X_{\rm sol})\delta^\mu_{~\nu}$.
This is equivalent to a cosmological constant $\Lambda=\beta{\cal L}(X_{\rm sol})$.

%======================================%
%<<<<<<<<<<<< SECTION I  >>>>>>>>>>>>>>%
%======================================%
\section{Jacobi fields and tidal forces in the spacetime (\ref{metric-app})}
\label{sec:tidal}

In this appendix, we study Jacobi fields and tidal forces in the spherically symmetric spacetime (\ref{metric-app}) along affinely parametrized ingoing radial timelike geodesics given by $x^\mu=(t(\lambda),r(\lambda),\theta_0,\phi_0)$, where $\theta_0$ and $\phi_0$ are constants.
For such a geodesic $\gamma$, $t(\lambda)$ and $r(\lambda)$ are determined by 
\begin{align}
{\dot t}=\frac{E}{f(r)},\qquad {\dot r}=-\sqrt{E^2-f(r)},\label{geo1}
\end{align}
which are Eqs.~(\ref{g-master2}) with $L=0$, $\varepsilon=-1$, and $C=0$.
Then, we introduce basis vectors $E^\mu_{(a)}~(a=0,1,2,3)$ in a parallelly propagated frame along $\gamma$ as
\begin{align}
E^\mu_{(0)}\frac{\partial}{\partial x^\mu}=&k^\mu\frac{\partial}{\partial x^\mu}=\frac{E}{f}\frac{\partial}{\partial t}+{\dot r}\frac{\partial}{\partial r},\\
E^\mu_{(1)}\frac{\partial}{\partial x^\mu}=&-\frac{{\dot r}}{f}\frac{\partial}{\partial t}-E\frac{\partial}{\partial r},\\
E^\mu_{(2)}\frac{\partial}{\partial x^\mu}=&\frac{1}{r}\frac{\partial}{\partial \theta},\qquad E^\mu_{(3)}\frac{\partial}{\partial x^\mu}=\frac{1}{r\sin\theta_0}\frac{\partial}{\partial \phi},
\end{align}
which satisfy $g_{\mu\nu}E^\mu_{(a)}E^\nu_{(b)}=\eta_{(a)(b)}={\rm diag}(-1,1,1,1)$ and $E^\nu_{(0)}\nabla_\nu E^\mu_{(a)}(={\dot E}^\mu_{(a)})=0$.

Using
\begin{align}
\begin{aligned}
R_{trtr}=&\frac12f'',\qquad R_{t\theta t\theta }=\frac{1}{\sin^2\theta}R_{t\phi t\phi}=\frac12rff',\\
R_{r\theta r\theta }=&\frac{1}{\sin^2\theta}R_{r\phi r\phi}=-\frac12rf^{-1}f',\qquad R_{\theta\phi\theta\phi}=r^2(1-f)\sin^2\theta,
\end{aligned}  
\end{align}  
we obtain non-zero orthonormal components of the Riemann tensor $R_{(a)(b)(c)(d)}:=R_{\mu\nu\rho\sigma}E^\mu_{(a)}E^\nu_{(b)}E^\rho_{(c)}E^\sigma_{(d)}$ along $\gamma$ as  
\begin{align}
R_{(0)(1)(0)(1)}=&\frac12f'',\label{R0101}\\
R_{(0)(2)(0)(2)}=&R_{(0)(3)(0)(3)}=\frac12r^{-1}f',\label{R0202}\\
R_{(1)(2)(1)(2)}=&R_{(1)(3)(1)(3)}=-\frac12r^{-1}f',\\
R_{(2)(3)(2)(3)}=&r^{-2}(1-f).\label{R2323}
\end{align}  
Jacobi fields $Z_{(I)}^{\mu}~(I=1,2,3)$ along $\gamma$ describe the spread of geodesics infinitesimally close to $\gamma$.
The geodesic deviation equations, or the Jacobi equations, to determine $Z_{(I)}^{\mu}$ are written as
\begin{align}
{\ddot Z}_{(I)}^\mu=&-{\cal K}^\mu_{~\rho}Z_{(I)}^{\rho},\label{jacobi}
\end{align}  
where ${\cal K}_{\mu\rho}$ is the tidal tensor defined by 
\begin{align}
{\cal K}_{\mu\rho}:=R_{\mu\nu\rho\sigma}k^\nu k^\sigma.
\end{align}  
Tidal forces along $\gamma$ are evaluated by the following orthonormal components of ${\cal K}_{\mu\rho}$:
\begin{align}
{\cal K}_{(a)(b)}={\cal K}_{\mu\rho}E^\mu_{(a)}E^\rho_{(b)}=R_{(a)(0)(b)(0)}.
\end{align}  
(See Sec.~7 in the textbook~\cite{Callahan}.)
Equations~(\ref{R0101}) and (\ref{R0202}) show the following non-zero components of ${\cal K}_{(a)(b)}$:
\begin{align}
{\cal K}_{(1)(1)}=\frac12f'', \qquad {\cal K}_{(2)(2)}={\cal K}_{(3)(3)}=\frac12r^{-1}f'.\label{K-tidal}
\end{align}  

If the mass function $M(r)$ in the metric (\ref{metric-app}) behaves around $r=0$ as $M(r)\simeq M_0 r^{3+\alpha}$ with $M_0\ne 0$, the metric function behaves as $f(r)\simeq 1-2M_0r^{2+\alpha}$.
Hereafter we assume $\alpha>-2$.
Then, since the radial geodesic equation~(\ref{geo1}) is integrated to give
\begin{align}
-(\lambda-\lambda_0)=\int^r\frac{\D {\bar r}}{\sqrt{E^2-f({\bar r})}},\label{radial-geo-sol}
\end{align}  
where $\lambda_0$ is an integration constant, a geodesic with $E^2> 1$ reaches $r=0$ with finite $\lambda$.
A geodesic with $E^2=1$ and $M_0>0$ reaches $r=0$ with finite $\lambda$ only for $-2<\alpha<0$ as Eq.~(\ref{radial-geo-sol}) gives
\begin{align}
\label{geodesic-r=0}
\lim_{r\to 0}\int^r\frac{\D {\bar r}}{\sqrt{E^2-f({\bar r})}}\simeq \left\{
\begin{array}{ll}
 -\frac{2}{\alpha\sqrt{2M_0}}r^{-\alpha/2} & (\alpha\ne 0)\\
\frac{1}{\sqrt{2M_0}}\ln|r| & (\alpha=0)
\end{array}
\right..
\end{align}
Thus, by Eqs.~(\ref{R0101})--(\ref{R2323}) and (\ref{K-tidal}), $R_{(a)(b)(c)(d)}$ and ${\cal K}_{(a)(b)}$ diverge as $r\to 0$ for $-2<\alpha< 0$ and remain finite for $\alpha\ge 0$ along a radial timelike geodesic $\gamma$ reaching $r=0$.

For $\alpha\ge 0$, Jacobi fields along $\gamma$ also remain finite as $r\to 0$.
For $-2<\alpha<0$, at least two Jacobi fields diverge as $r\to 0$ along $\gamma$ with $E^2=1$.
To show this, we write Jacobi fields along ${\gamma}$ as
\begin{align}
Z^\mu_{(I)}\frac{\partial}{\partial x^\mu}=l_{(I)} E_{(I)}^\mu\frac{\partial}{\partial x^\mu},\label{Z-l}
\end{align}  
where $l_{(I)}=l_{(I)}(\lambda)$ describe norms of the Jacobi fields as $g_{\mu\nu}Z^\mu_{(I)}Z^\nu_{(I)}=l_{(I)}^2$.
Substituting Eq.~(\ref{Z-l}) into the Jacobi equations (\ref{jacobi}) and using ${\dot E}^\mu_{(I)}=0$, we obtain
\begin{align}
{\ddot l}_{(I)}=- R_{(I)(0)(I)(0)}l_{(I)}, \label{l-eq0}
\end{align}  
which gives
\begin{align}
&2(E^2-f)l_{(1)}''-f' l_{(1)}'+f''l_{(1)}=0,\label{ode-l1}\\
&2(E^2-f)l_{(I)}''-f' l_{(I)}'+r^{-1}f'l_{(I)}=0~~(\mbox{for}~I=2,3) \label{ode-l2}
\end{align}
for $l_{(I)}=l_{(I)}(\lambda(r))~(I=1,2,3)$, where we used Eqs.~(\ref{geo1}), (\ref{R0101}), and (\ref{R0202}).
The general solutions to Eqs.~(\ref{ode-l1}) and (\ref{ode-l2}) are respectively given by 
\begin{align}
l_{(1)}=&C_1\sqrt{E^2-f(r)}\biggl(D_1+\int^r\frac{\D {\bar r}}{(E^2-f({\bar r}))^{3/2}}\biggl),\label{l1}\\
l_{(I)}=&-C_Ir\biggl(D_I+\int^r\frac{\D {\bar r}}{{\bar r}^2\sqrt{E^2-f({\bar r})}}\biggl)~~(\mbox{for}~I=2,3),\label{l23}
\end{align}  
where $C_I$ and $D_I~(I=1,2,3)$ are integration constants.
Equations~(\ref{l1}) and (\ref{l23}) show that $l_{(I)}$ (and hence $Z^\mu_{(I)}$) remain finite as $r\to 0$ along $\gamma$ with $E^2\ne 1$ for $\alpha>-2$.
Along $\gamma$ with $E^2=1$ for $-2<\alpha<0$ and $M_0>0$, we obtain
\begin{align}
\lim_{r\to 0}\int^r\frac{\D {\bar r}}{(E^2-f({\bar r}))^{3/2}}\simeq& \left\{
\begin{array}{ll}
-\frac{2}{(4+3\alpha)(2M_0)^{3/2}}r^{-(4+3\alpha)/2} & (\alpha\ne -4/3)\\
\frac{1}{(2M_0)^{3/2}}\ln|r| & (\alpha= -4/3)
\end{array}
\right.,\\
\lim_{r\to 0}\int^r\frac{\D {\bar r}}{{\bar r}^2\sqrt{E^2-f({\bar r})}}\simeq&-\frac{2}{\sqrt{2M_0}(4+\alpha)r^{(4+\alpha)/2}}~~(\mbox{for}~I=2,3),
\end{align}  
so that Eqs.~(\ref{l1}) and (\ref{l23}) show that $l_{(1)}$ and $l_{(I)}~(I=2,3)$ diverge as $r\to 0$ for $-1<\alpha<0$ and $-2<\alpha<0$, respectively.

\end{document}